\documentclass[11pt]{article} 
\pdfoutput=1
\usepackage{amsmath,amssymb,amsfonts,fullpage,graphicx,multirow,nicefrac,slashed}
\usepackage[compress]{cite}
\usepackage[normalem]{ulem} 
\usepackage[normal,font=small,labelfont=bf,labelsep=period]{caption}
\renewcommand{\arraystretch}{1.2}
\renewcommand{\hat}{\widehat}

\newcommand{\br}{\mathcal{BR}}

\usepackage{color}
\definecolor{colorLink}{rgb}{0.9,0,0} 
\definecolor{colorCite}{rgb}{0,0.7,0} 
\definecolor{colorURL} {rgb}{0,0,0.8} 
\definecolor{colorCapt}{rgb}{0.5,0,1} 
\definecolor{colorPink}{rgb}{1,0,0.7} 
\usepackage[colorlinks=true,linktocpage=true,linkcolor=colorLink,citecolor=colorCite,urlcolor=colorURL]{hyperref}

\title{\vspace{-2cm}
\begin{flushright}{\small EFI-15-40}\vspace{1cm}\end{flushright}
\noindent{\LARGE \bf A pseudoscalar decaying to photon pairs \\ in the early LHC Run 2 data}\\}

\author{Matthew Low$^a$, Andrea Tesi$^b$, Lian-Tao Wang$^c$}
\date{\it \small $^a$School of Natural Sciences, Institute for Advanced Study, Princeton, NJ 08540 \\ \vspace{0.8em}
$^b$Department of Physics, Enrico Fermi Institute, University of Chicago, Chicago, IL 60637\\ \vspace{0.8em}
$^c$Department of Physics, Enrico Fermi Institute, and Kavli Institute for\\
Cosmological Physics, University of Chicago, Chicago, IL 60637}

\begin{document}
\begin{titlepage}
\maketitle
\thispagestyle{empty}

\begin{abstract}
\noindent
In this paper we explore the possibility of a pseudoscalar resonance to account for the 750 GeV diphoton excess observed both at ATLAS and at CMS.  We analyze the ingredients needed from the low energy perspective to obtain a sufficiently large diphoton rate to explain the signal while avoiding constraints from other channels.  Additionally, we point out composite Higgs models in which one can naturally obtain a pseudoscalar at the 750 GeV mass scale and we estimate the pseudoscalar couplings to standard model particles that one would have in such models.  A generic feature of models that can explain the excess is the presence of new particles in addition to the 750 GeV state.  Finally, we note that due to the origin of the coupling of the resonance to photons, one expects to see comparable signals in the $Z\gamma$, $ZZ$, and $WW$ channels.
\end{abstract}

\vfill
\noindent\line(1,0){188}\\\medskip
\footnotesize{E-mail: \texttt{\href{mailto:mattlow@ias.edu}{mattlow@ias.edu}, \href{mailtoatesi@uchicago.edu}{atesi@uchicago.edu}, \href{mailto:liantaow@uchicago.edu}{liantaow@uchicago.edu}}}
\end{titlepage}
\tableofcontents

\section{Motivation}\label{sec:intro}

With the start of the second run of the Large Hadron Collider (LHC), we are seeing the first glimpses into physics at collision energies of 13 TeV.  So far ATLAS and CMS have only collected a small amount of data (3.2 fb$^{-1}$ and 2.6 fb$^{-1}$ respectively), but that is already enough to set competitive limits on certain classes of new particles.  For instance, jets and missing energy searches are already setting stronger limits on gluinos than at 8 TeV, due to the quickly growing parton luminosities at high masses.  For new particles at lower masses, however, the parton luminosity increase is much milder and in most cases the 13 TeV searches have not yet surpassed the 8 TeV searches in sensitivity.

One 13 TeV search that has received significant attention recently is the diphoton resonance search.  Both ATLAS~\cite{ATLAS-CONF-2015-081} and CMS~\cite{CMS-PAS-EXO-15-004} observe an excess at 750 GeV.  It appears that the excess is compatible both with Run 1 data and between ATLAS and CMS which makes this a compelling case of potential new physics.  In this paper we explore the model building possibilities to describe this excess from two complementary perspectives.  The first perspective we take is to quantitatively analyze the low energy interactions needed to produce the observed diphoton rate.  For this we identify the diphoton resonance as a new pseudoscalar particle that couples to the standard model (SM) through dimension 5 operators.  The generic picture is that the pseudoscalar is produced in gluon fusion and then decays to a pair of photons.  We show that this parametrization can account for the excess either with a large enhancement in the coupling to photons or with a moderate enhancement in the coupling to photons and gluons and moderate suppression in the couplings to fermions.  Either of these cases implies that more particles in addition to the 750 GeV resonance are needed to fit the data.

The second aspect of the diphoton excess that we address is naturally finding a scalar (or pseudoscalar) with a mass of $\sim$750 GeV in a complete model.  We know from the familiar example of the Higgs that theories with fundamental scalars appearing much below the cutoff are finetuned.  One way this finetuning problem has been addressed is to posit that the Higgs is actually a composite particle of some new strong dynamics.  While this idea solves the hierarchy problem in principle, in practice there is still residual tuning associated with a light Higgs meaning that we are forced to live with some level of tuning.  It could be case, however, that there are other scalars coming from the strong dynamics that are not tuned.  In other words, it could be that the Higgs as a pseudo Nambu Goldstone Boson (pNGB) is slightly tuned, but that the other pNGBs are at their naturalness limit.  We will argue that in the composite Higgs framework one can have additional light scalars at the 750 GeV mass scale and that such (pseudo)scalars are compatible with the excess.

Given the minimal information about the diphoton resonance, one cannot conclusively associate the resonance to the pseudoscalar parametrization that we present.  We therefore survey a few other model building possibilities along with a few simple estimates to assess how easily these alternative models can fit the excess in comparison to the pseudoscalar case.  In particular we look at a scalar resonance and a spin-2 resonance.

The outline is as follows.  In Sec.~\ref{sec:lhc} we review the current experiment status of the diphoton resonance and collect limits from other potentially relevant channels.  The interactions of a pseudoscalar are described in Sec.~\ref{sec:interactions} along with computations of widths, branching ratios, and rates.  In Sec.~\ref{sec:pseudoscalarmass} we address the issue of getting the $\sim$750 GeV mass scale in composite Higgs models.  To conclude, in Sec.~\ref{sec:alternatives} we point out other possibilities and summarize in Sec.~\ref{sec:conclusions}.  App.~\ref{sec:app} provides details of the SO(6)/SO(5) composite Higgs model that contains a pseudoscalar pNGB.

\section{Signals and constraints from the LHC}\label{sec:lhc}

ATLAS and CMS have both reported excess in the diphoton channel at a mass very near to 750 GeV.  For a narrow resonance, the local significance reported by ATLAS was 3.6$\sigma$ and 2.6$\sigma$ by CMS.  When a wide resonance signal model is used, the significances shift to 3.9$\sigma$ for ATLAS and 2.0$\sigma$ for CMS.  In both experiments the global significance is $\sim 2\sigma$.

To gain some idea of the expected sensitivity, we compile the expected and observed limits set by Run 1 diphoton searches in Table~\ref{tab:diphoton8tev}.

\begin{table} [h]
\begin{center}
\begin{tabular}{c||ccc}
$\gamma\gamma$     & expected & observed & \\ \hline\hline
ATLAS (spin-2)     &   1.9 fb &   2.4 fb & ~\cite{Aad:2015mna} \\
CMS (spin-2)       &   1.5 fb &   1.9 fb & ~\cite{CMS:2015cwa} \\
CMS (narrow)       &   0.7 fb &   1.3 fb & ~\cite{Khachatryan:2015qba} \\
CMS (wide)         &   2.0 fb &   2.3 fb & ~\cite{Khachatryan:2015qba}
\end{tabular}
\caption{Upper limits (at 95\% CL) on the $\sigma \times \br$ of a 750 GeV resonance decaying to a pair of photons from 8 TeV LHC data.}
\label{tab:diphoton8tev}
\end{center}
\end{table}

While a proper analysis should perform a combination of both the 8 TeV and 13 TeV results from both experiments to assess the compatibility of the signal and the correct cross section to fit, this is difficult to do reliably with such a small number of events.  As such we will show the cross sections that can be obtained with a pseudoscalar resonance rather than fixing a signal strength value.  As a guide, one can use the CMS combination of their 8 TeV and 13 TeV results which finds a cross section of $\sim 3-5$ fb~\cite{slides}.  

\begin{table} [t]
\begin{center}
\begin{tabular}{cc||cc}
final state          &              & observed & \\ \hline \hline
$t\bar{t}$           & scalar       &  700 fb  & ATLAS~\cite{Aad:2015fna}       \\
$t\bar{t}$           & spin-2       &  540 fb  & ATLAS~\cite{Aad:2015fna}       \\
$t\bar{t}$           & narrow       &  450 fb  & CMS~\cite{Khachatryan:2015sma} \\
$t\bar{t}$           & wide         &  510 fb  & CMS~\cite{Khachatryan:2015sma} \\ \hline
$b\bar{b}$           &              &  1.2 pb  & CMS~\cite{Khachatryan:2015tra} \\ \hline
$Z\gamma$            &              &  2.7 fb  & ATLAS~\cite{Aad:2014fha}       \\ \hline
$ZZ$                 & scalar       &   12 fb  & ATLAS~\cite{Aad:2015kna}       \\ 
$ZZ$                 & spin-2       &   38 fb  & ATLAS~\cite{Aad:2014xka}       \\ 
$ZZ$                 & scalar       &   23 fb  & CMS~\cite{Khachatryan:2015cwa} \\
$ZZ$                 & spin-2       &   53 fb  & CMS~\cite{Khachatryan:2014gha} \\ \hline
$WW$                 & spin-2       &   67 fb  & ATLAS~\cite{Aad:2015ufa}       \\ 
$WW$                 & scalar       &   47 fb  & CMS~\cite{Khachatryan:2015cwa} \\ \hline
$jj$                 & Gaussian     &  2.0 pb  & ATLAS~\cite{Aad:2014aqa}       \\ 
$jj$                 & Breit Wigner & 20.0 pb  & ATLAS~\cite{Aad:2014aqa}       \\
$jj$                 &              &  2.9 pb  & CMS~\cite{CMS-PAS-EXO-14-005}  \\ \hline
$\ell^+\ell^-$       & spin-2       &  1.1 fb  & ATLAS~\cite{Aad:2014cka}       \\
$\ell^+\ell^-$       & spin-2       &  3.5 fb  & CMS~\cite{Khachatryan:2014fba} \\ \hline
$hh$                 &              &   32 fb  & ATLAS~\cite{Aad:2015uka}       \\
$hh$                 & scalar       &   51 fb  & CMS~\cite{Khachatryan:2015yea} \\
$hh$                 & spin-2       &   39 fb  & CMS~\cite{Khachatryan:2015yea}
\end{tabular}
\caption{Observed upper limits (at 95\% CL) on $\sigma \times \br$ of a 750 GeV resonance decaying to various final states from 8 TeV LHC data.}
\label{tab:limits8tev}
\end{center}
\end{table}

In Table~\ref{tab:limits8tev} we list the observed limits from other channels that can be applicable to models that explain the diphoton excess.  The limits shown are the observed limits and are set on $\sigma \times \br$.  For dijet limits we use the reported acceptance of $\mathcal{A}=0.6$ for spin-0 signals to cast the limit from $\sigma \times \br \times \mathcal{A}$ to $\sigma \times \br$~\cite{Aad:2014aqa,CMS-PAS-EXO-14-005}.  There are also searches for resonances in the $\tau^+\tau^-$~\cite{Khachatryan:2014wca,Aad:2014vgg}, $Zh$~\cite{Aad:2015wra}, and monojet~\cite{Khachatryan:2014rra,Aad:2015zva} channels which can be relevant for particular models.

\begin{table} [h]
\renewcommand{\arraystretch}{1.2}
\begin{center}
\begin{tabular}{c||cc}
final state          & scaled    \\ \hline \hline
$t\bar{t}$           &  2.1 pb   \\
$b\bar{b}$           &  5.6 pb   \\
$Z\gamma$            &   13 fb   \\
$ZZ$                 &   56 fb   \\
$WW$                 &  220 fb   \\
$jj$                 &  9.4 pb   \\
$\ell^+\ell^-$       &  5.2 fb   \\
$hh$                 &  150 fb  
\end{tabular}
\caption{Observed LHC limits at 13 TeV on $\sigma \times \br$  rescaled from 8 TeV using the $gg$ parton luminosity~\cite{partonlumi}.}
\label{tab:limits13tev}
\end{center}
\end{table}

In Table~\ref{tab:limits13tev} we rescale the strongest 8 TeV limits by their $gg$ parton luminosity ratio~\cite{partonlumi} because in the models we consider the production is dominated by gluon fusion.  A strict comparison of compatibility of a proposed model with 8 TeV limits would involve simulating the signal model at 8 TeV but the numbers in Table~\ref{tab:limits13tev} offer a quick comparison.

Finally, we note that the observed signal rate of $\sim 3-5$ fb is rather large.  In the case of the SM Higgs, the decays to photons are mediated by loops of tops and $W$'s and lead to a diphoton branching ratio of $\sim 10^{-3}$.  If the decays of the 750 GeV resonance to photons were likewise only mediated by tops and $W$'s the diphoton ratio would be small, $\lesssim 10^{-4}$ (because $WW$ and $ZZ$ decays are now onshell), which would require rates to $t\bar{t}$ and $WW$ of $\mathcal{O}({\rm pb})$ at Run 1.  From Table~\ref{tab:limits8tev} this is clearly ruled out.  Thus one can conclude that for a sufficiently large diphoton rate the 750 GeV is not the only new particle, more are needed!

\section{The interactions of a pseudoscalar}\label{sec:interactions}

A spin-0 particle can either be a scalar or a pseudoscalar.  The simplest possibility to start with is to consider an SM singlet.  A scalar singlet can potentially mix with the Higgs which would introduce tree level decays to $t\bar{t}$, $WW$, $ZZ$ and even $hh$, which can place strong constraints on the mixing.  It also suppresses the rate to photons compared to $VV$ similarly to the case of a heavy SM Higgs of mass 750 GeV.  This very fact together with the relative importance of the diboson channels (see Table~\ref{tab:limits13tev}) requires a huge contribution to the diphoton rate from new physics or a tuning of the mixing.  Assuming CP conservation, a pseudoscalar will not mix with the Higgs which makes explaining the excess easier.\footnote{A scalar as part of an additional doublet is another scenario that can be safe from mixing with the Higgs.}  We will therefore focus our discussion on a pseudoscalar resonance, and reserve comments on the scalar case until Sec.~\ref{sec:alternatives}.

We consider the SM extended by the addition of an SM singlet pseudoscalar $\eta$ which transforms under CP as
\begin{equation}
\eta \xrightarrow{\rm CP} -\eta.
\end{equation}
The scalar potential is given by
\begin{equation} \label{eq:potential}
V = V_{\rm SM} + \frac{m_\eta^2}{2} \eta^2 + \frac{\lambda_\eta}{4!} \eta^4 + \frac{\lambda_{\eta h}}{2} \eta^2 |H|^2.
\end{equation}
We assume that CP is conserved, which at the level of the scalar potential simply acts as a $Z_2$ symmetry on $\eta$.  This forbids mixing with the Higgs.  The difference between $Z_2$ and CP becomes apparent when one considers non-renormalizable interactions.  At dimension 5 the only interactions involving $\eta$ are
\begin{equation} \label{eq:interactions}
\mathcal{L}_{\rm int} = 
\frac{y_f}{\Lambda_f} \eta ( i \overline{f_L}H f_R + {\rm h.c.})
+ \frac{c_B}{\Lambda_g} \frac{g'^2}{16\pi^2} \eta B_{\mu\nu} \tilde{B}^{\mu\nu}
+ \frac{c_W}{\Lambda_g} \frac{g^2}{16\pi^2} \eta W^a_{\mu\nu} \tilde{W}^{a\mu\nu}
+ \frac{c_g}{\Lambda_g} \frac{\alpha_s}{4\pi} \eta G^a_{\mu\nu} \tilde{G}^{a\mu\nu} ,
\end{equation}
where $y_f$ is the Yukawa coupling of the fermion $f$ and $c_B$, $c_W$, and $c_g$ are parameters.  For simplicity we supress all fermion operators by a common scale $\Lambda_f$ and all gauge field operators by a common scale $\Lambda_g$.  These scales can of course be different and it is straightforward to generalize Eq.~\eqref{eq:interactions}.\footnote{Although one should note that in our parametrization the difference in gauge scales can be absorbed into $c_B$, $c_W$, and $c_g$.}  The normalization we use is $\tilde{B}^{\mu\nu} = \epsilon^{\mu\nu\alpha\beta} B_{\alpha\beta}$.

Notice that loops of SM fermions will already contribute to the interactions between the pseudoscalar and gauge boson pairs.  The parameters $c_B$, $c_W$, and $c_g$ in Eq.~\eqref{eq:interactions} denote contributions in addition to those from SM loops.   As we are particularly interested in the diphoton rate, we define the parameter $c_\gamma = c_B + c_W$ which denotes the additional UV contribution to $\eta F_{\mu\nu} \tilde{F}^{\mu\nu}$.

One possibility for UV physics that could generate the dimension 5 operators above are heavy vector-like particles.  In this case, one needs to be careful that the new particles do not lead to additional signals that would rule out the pseudoscalar explanation.  For instance, requiring the new particles to be heavier than half the pseudoscalar mass protects against large branching ratios to these new particles.  The limit where the new particles are just above threshold is interesting as the loop functions are maximal at threshold and could lead to sizable effects.  In this work, these effects are parametrized with the aforementioned operator coefficients.

After fixing the pseudoscalar mass to 750 GeV, the parameter space consists of two dimensionful parameters $\Lambda_f$ and $\Lambda_g$ and three dimensionless parameters $c_B$, $c_W$, and $c_g$.  One can see from the fact that $c_\gamma = c_B + c_W$ that the diphoton coupling can be increased by enhancing either the $\eta B_{\mu\nu} \tilde{B}^{\mu\nu}$ operators or the $\eta W^a_{\mu\nu} \tilde{W}^{a\mu\nu}$ operator.  Increasing $c_W$ will increase the $WW$ coupling as well.  In this work we set $c_W=0$ for simplicity such that branching ratio to $WW$ vanishes and $WW$ resonance searches are not constraining.  The parameter space is  $(\Lambda_f, \Lambda_g, c_\gamma, c_g)$.

\begin{figure}[t]
\begin{center}
\includegraphics[width=.45\textwidth]{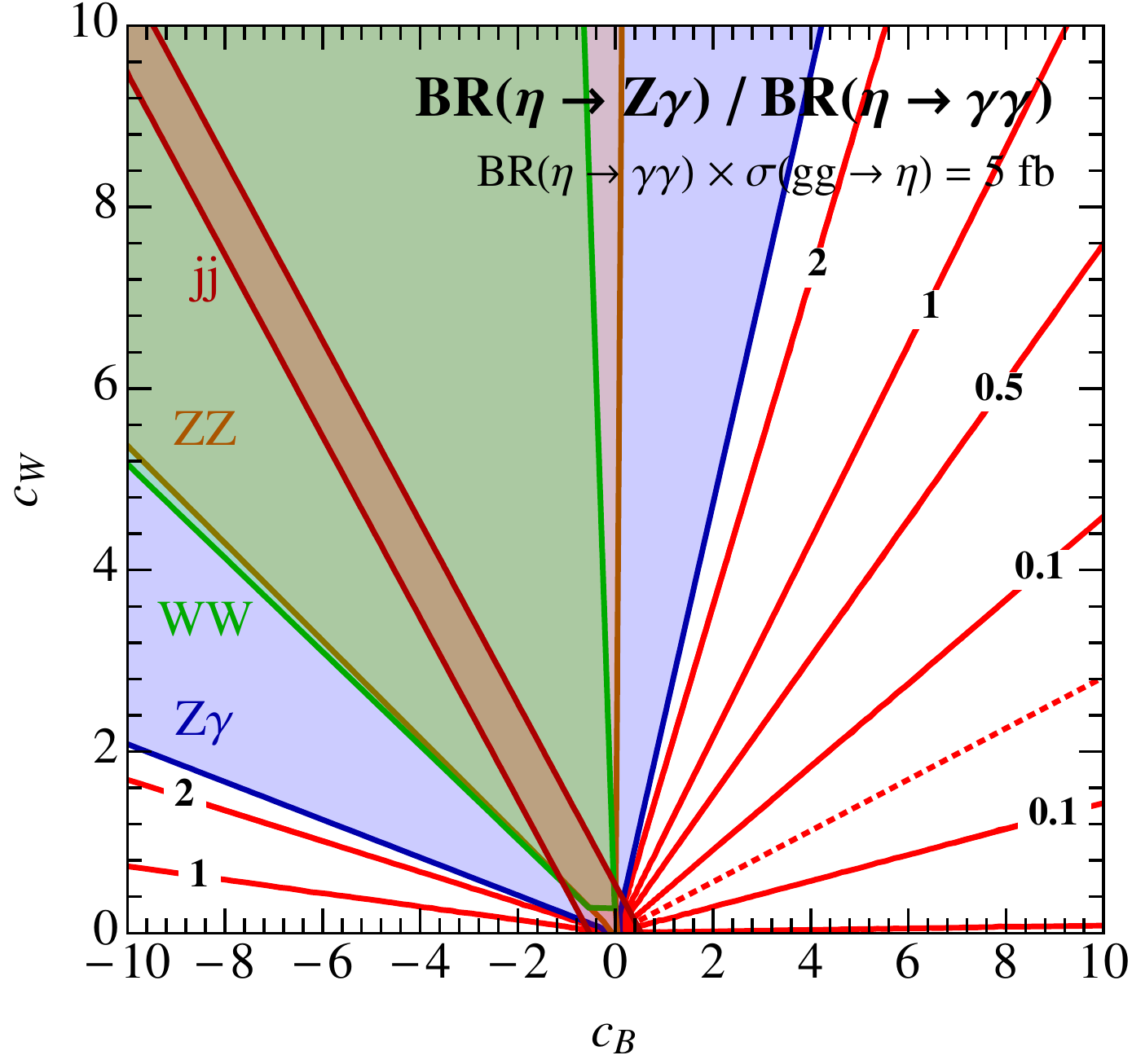} \quad\quad
\includegraphics[width=.45\textwidth]{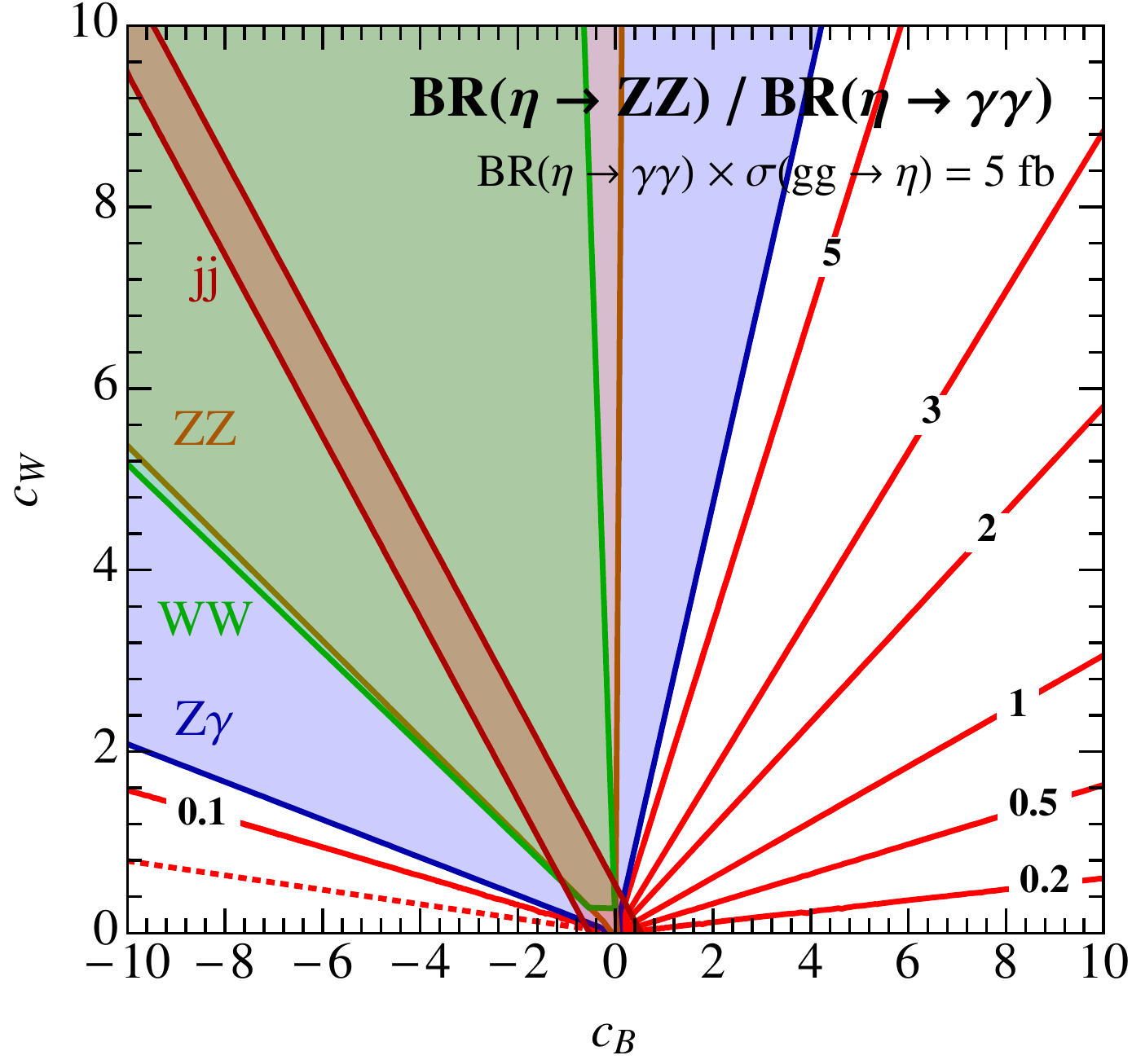}
\caption{Branching ratio to $Z\gamma$ (left) and $ZZ$ (right) normalized to the diphoton branching ratio.  The signal rate is fixed to 5 fb and the colored regions are excluded by 8 TeV diboson searches.  The dotted red line shows where the branching ratio vanishes.}
\label{fig:ratios}
\end{center}
\end{figure}

It is also interesting to study the case where $c_W \neq 0$.  When this is the case, $WW$ resonances searches become constraining in addition to constraints already from $Z\gamma$ and $ZZ$.  While one can select combinations of $c_B$ and $c_W$ to set any of the branching ratios to $Z\gamma$, $ZZ$, or $WW$, to zero, the other two are necessarily non zero.  In this sense, a generic prediction of the diphoton signal is a signal in two or more of the corresponding diboson channels.  In Fig.~\ref{fig:ratios} we show the branching ratio to $Z\gamma$ (left) and $ZZ$ (right) normalized to the diphoton branching ratio.  In the plot $\Lambda_f$ is decoupled and consequently $t\bar{t}$ searches are not relevant.

\subsection*{Partial Widths}

Given the interactions in Eq.~\eqref{eq:interactions} we can compute the partial decay widths.  We only show the most relevant which are $t\bar{t}$, $gg$, $\gamma\gamma$, and to a lesser extent, $b\bar{b}$.
\begin{subequations}\label{eq:BRs}\begin{align}
\Gamma_{t\bar{t}} &=
   \frac{N_c}{8\pi} \frac{m_t^2}{\Lambda_f^2} m_\eta \sqrt{1-\frac{4m_t^2}{m_\eta^2}}, \\
\Gamma_{b\bar{b}} &=
   \frac{N_c}{8\pi} \frac{m_b^2}{\Lambda_f^2} m_\eta \sqrt{1-\frac{4m_b^2}{m_\eta^2}}, \\
\Gamma_{gg} &=
   \frac{1}{2\pi} \left(\frac{\alpha_s}{4\pi}\right)^2 \frac{m_\eta^3}{\Lambda_f^2} \left| A_{-}(\tau) + 2 c_g \frac{\Lambda_f}{\Lambda_g}\right|^2, \\
\Gamma_{\gamma\gamma} &=
   \frac{1}{4\pi} \left(\frac{\alpha}{4\pi}\right)^2 \frac{m_\eta^3}{\Lambda_f^2} \left| N_c Q_t^2 A_{-}(\tau) + 2 c_\gamma \frac{\Lambda_f}{\Lambda_g} \right|^2 ,
\end{align}\end{subequations}
where $A_{-}(\tau)$ is the pseudoscalar loop function
\begin{equation}
A_{-}(\tau)= \tau f(\tau),
\quad\quad\quad
\tau = \frac{4m_f^2}{m_\eta^2},
\end{equation}
and the function $f(\tau)$ is given by
\begin{equation}
f(\tau)= \theta(\tau-1) \arcsin^2\left(\frac{1}{\sqrt{\tau}}\right)
       - \theta(1-\tau) \frac{1}{4} \left( \log \frac{1+\sqrt{1-\tau}}{1-\sqrt{1-\tau}} -i\pi \right)^2 .
\end{equation}
The branching ratios to $Z\gamma$, $WW$, and $ZZ$ are correlated due to SU(2) gauge invariance.  In the limit where we neglect the top loop (which is appropriate in the relevant parameter space) and $c_W=0$ the ratios are\footnote{Recall that this only holds when $c_W=0$.}
\begin{equation}
\br(\gamma\gamma) : \br(Z\gamma) : \br(ZZ) : \br(WW)
= 1 : 2 t_w^2 : t_w^4 : 0 .
\end{equation}
From Table~\ref{tab:limits13tev} one can see that for the appropriate diphoton signal, none of the diboson channels are constraining.

\begin{figure}[t]
\begin{center}
\includegraphics[width=.45\textwidth]{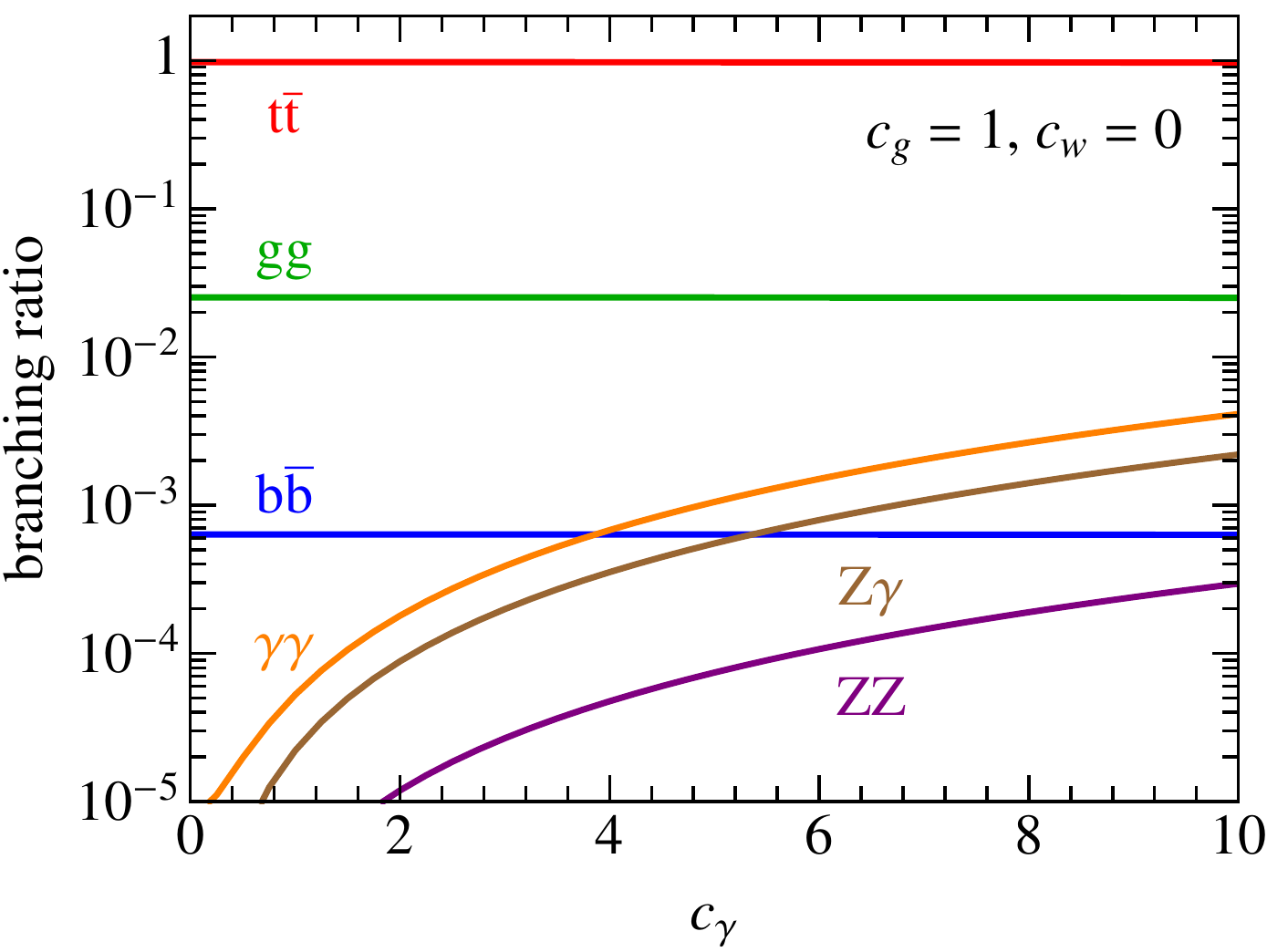} \quad\quad
\includegraphics[width=.45\textwidth]{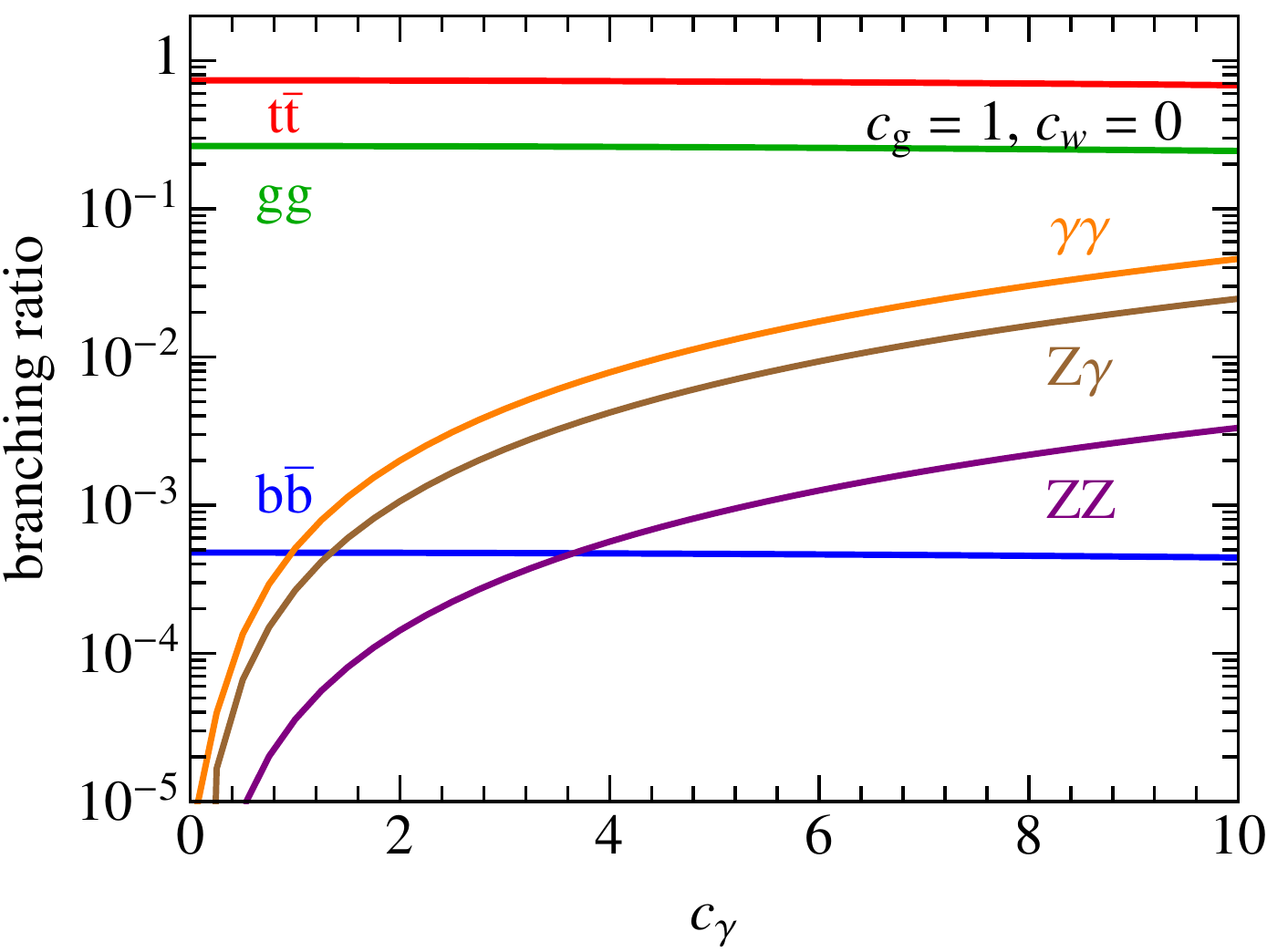}
\caption{Branching ratios of the pseudoscalar as a function of $c_\gamma$ which parameterizes UV contributions to the pseudoscalar-photon-photon interactions.  The parameters used are $\Lambda_g=$ 500 GeV and $\Lambda_f=$ 750 GeV (left) and $\Lambda_f=$ 3 TeV (right).}
\label{fig:branchingratios}
\end{center}
\end{figure}

In Fig.~\ref{fig:branchingratios} we show the branching ratios as a function of $c_\gamma$ for two values of $\Lambda_f =$ 750 GeV and $\Lambda_f=$ 3 TeV while $\Lambda_g =$ 500 GeV and $c_g=1$.  We see that $t\bar{t}$ dominates the branching ratio unless it is supressed by a very large $\Lambda_f$ value.  The branching ratios to $Z\gamma$ and $ZZ$ are estimated by only including their contribution from $c_\gamma$ and neglecting the top loop contribution to their partial widths.

From Eq.~\eqref{eq:BRs} one can quickly estimate the width to be
\begin{equation} \label{eq:width-estimate}
\frac{\Gamma}{m_\eta} \simeq \frac{N_c}{8\pi}\frac{m_t^2}{\Lambda_f^2} 
\simeq 10^{-2}\left(\frac{600~\mathrm{GeV}}{\Lambda_f}\right)^2 .
\end{equation}
The pseudoscalar tends to be narrow especially when $\Lambda_f$ becomes very large.

\subsection*{Production rate}

\begin{figure}[t]
\begin{center}
\includegraphics[width=.45\textwidth]{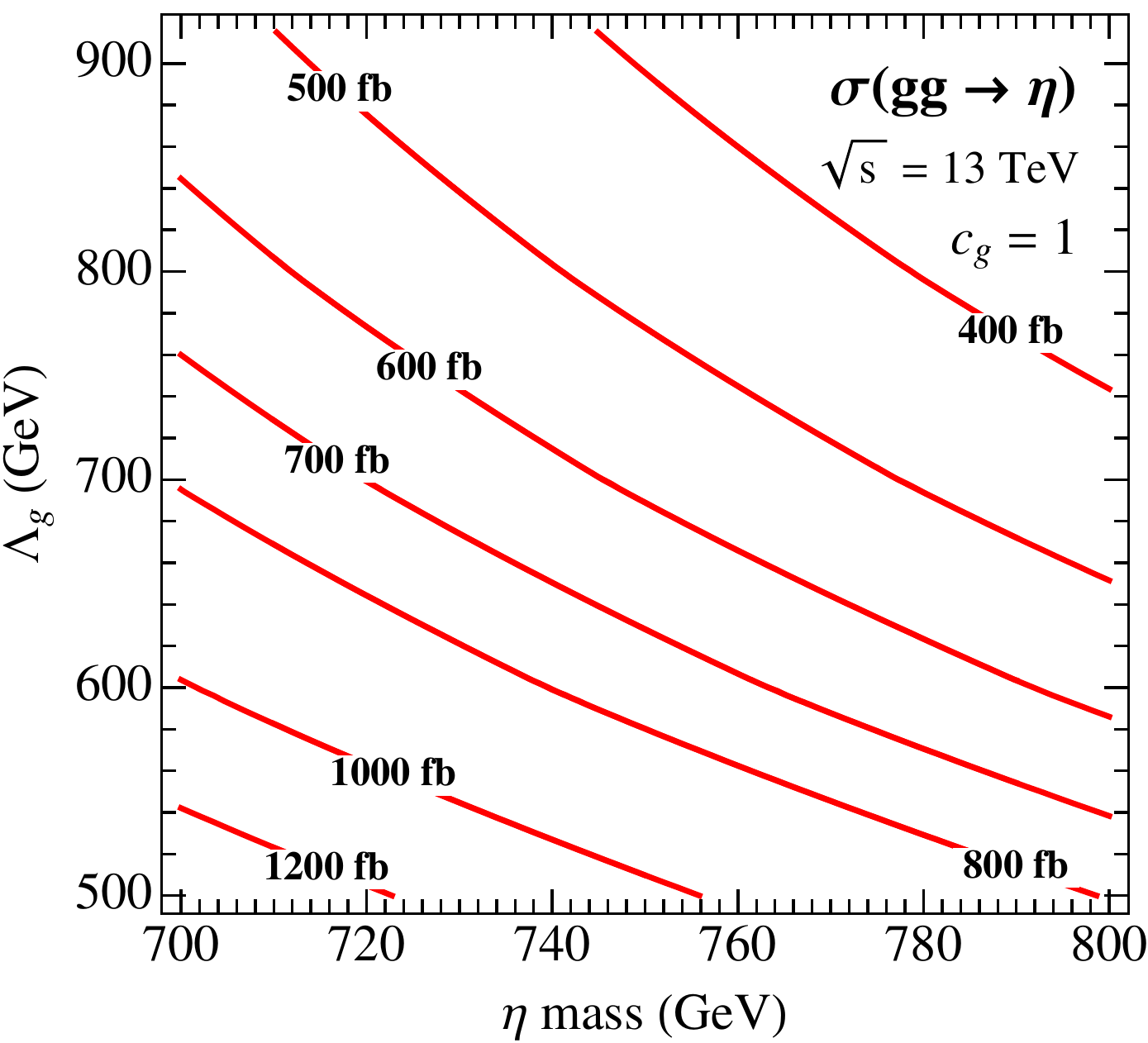} \quad\quad
\includegraphics[width=.45\textwidth]{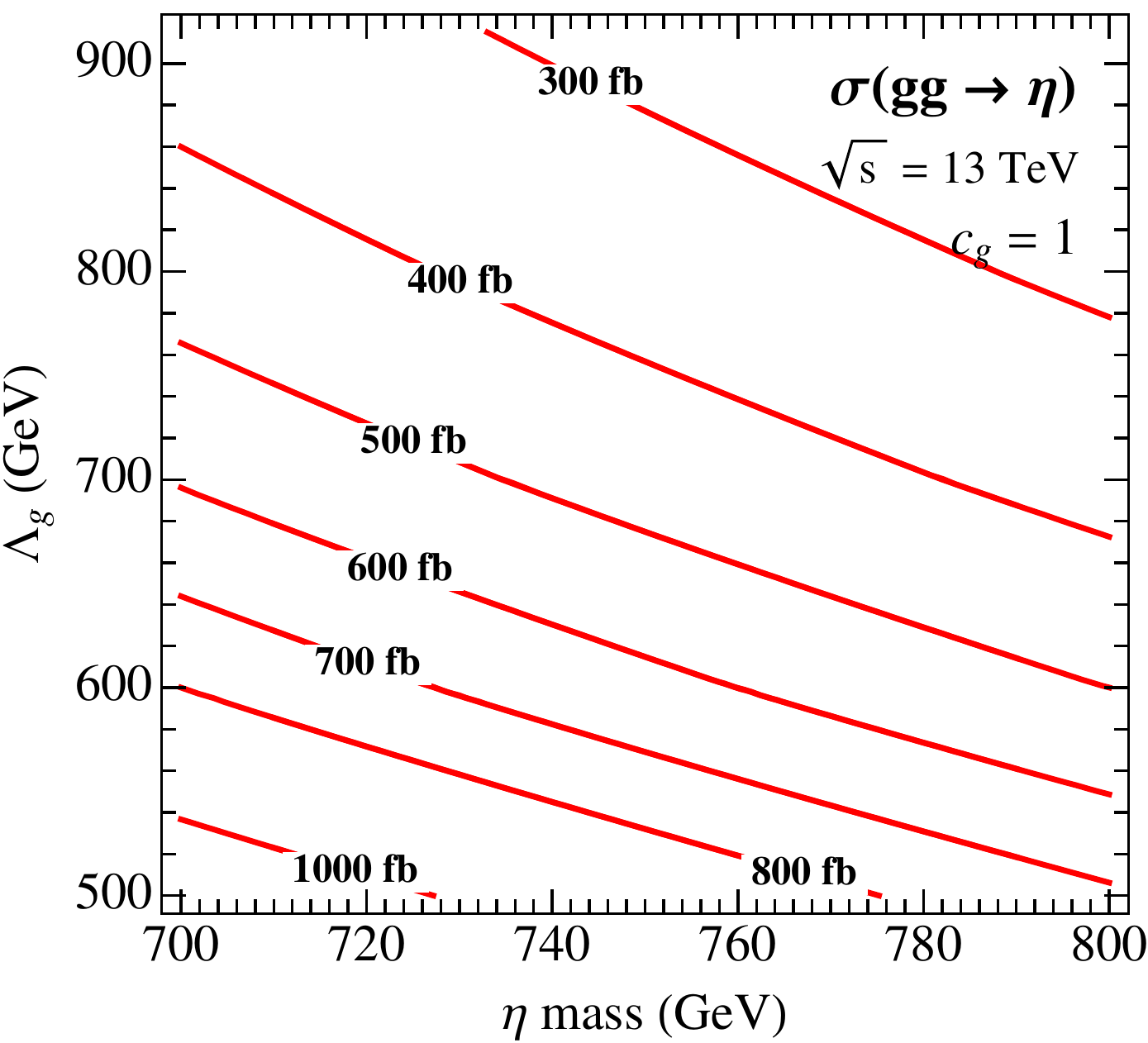}
\caption{The production cross section of the pseudoscalar as a function of the mass and $\Lambda_g$ using $\Lambda_f=$ 750 GeV (left) and $\Lambda_f=$ 3 TeV (right).}
\label{fig:production}
\end{center}
\end{figure}

From the branching ratios, one can see that the $\eta$ is produced in gluon fusion.  We show the total production cross section as a function of mass and the gauge field scale $\Lambda_g$ in Fig.~\ref{fig:production} for $\Lambda_f=$ 750 GeV (left) and $\Lambda_f=$ 3 TeV (right).  The only SM fermion we include in the loop is the top quark.  We compute the pseudoscalar cross section at NNLO using HIGLU~\cite{Spira:1995mt} and rescale the cross section to account for an additional gluon fusion contribution via $c_g = 1$.  The value of $\Lambda_g$ controls the relative rate due to the additional dimension 5 contribution.

Simple estimates of the production rate are useful and straightforward to obtain using information provided by the Higgs working group~\cite{Heinemeyer:2013tqa} which provides the production rates for heavy Higgses produced in gluon fusion as a function of mass at 8 TeV.  First, one needs to account for the difference between scalar and pseudoscalar production.  At leading order difference can be obtained by the ratio of loop functions
\begin{equation} \label{eq:xsecrescaling}
\sigma_\eta = \left|\frac{3}{2}\frac{A_{-}(\nicefrac{4m_t^2}{m_\eta^2})}{A_{+}(\nicefrac{4m_t^2}{m_\eta^2})}\right|^2 \sigma_H,
\end{equation}
where $A_{+}(\tau)$ is the scalar loop function
\begin{equation}
A_{+}(\tau) = \frac{3}{2}\tau(1+(1-\tau)f(\tau)).
\end{equation}
At 750 GeV this ratio works out to be $\simeq$ 1.45.  Next, one can rescale the 8 TeV rates to 13 TeV by the parton luminosities which is 4.7 for a $gg$ initial state~\cite{partonlumi}.  Finally one needs to account for the prefactor of the pseudoscalar-top coupling in Eq.~\eqref{eq:interactions} relative to the Higgs-top coupling in the standard model.  Compiling these numbers together and rescaling from the NNLL QCD + NLO electroweak 8 TeV rate, one finds the rate at 13 TeV to be (for $c_g= 0$)
\begin{equation}
\sigma_\eta(750~\mathrm{GeV})\bigg|_{c_g=0} = \left(\frac{v}{\Lambda_f}\right)^2 \times 1.0~\mathrm{pb}.
\end{equation}
Therefore given a mild suppression from $(v/\Lambda_f)^2$ and a diphoton branching ratio of $\sim 10^{-3} - 10^{-2}$ one can see that the diphoton rate will be $\mathcal{O}(5~\rm fb)$ as is needed to explain the excess.  Allowing for a non vanishing $c_g$ the above result is then rescaled (and typically enhanced) by the ratio of the partial width to gluons in the two cases, $\Gamma_{gg}(c_g)/\Gamma_{gg}(c_g=0)$.

\begin{figure}[t]
\begin{center}
\includegraphics[width=.45\textwidth]{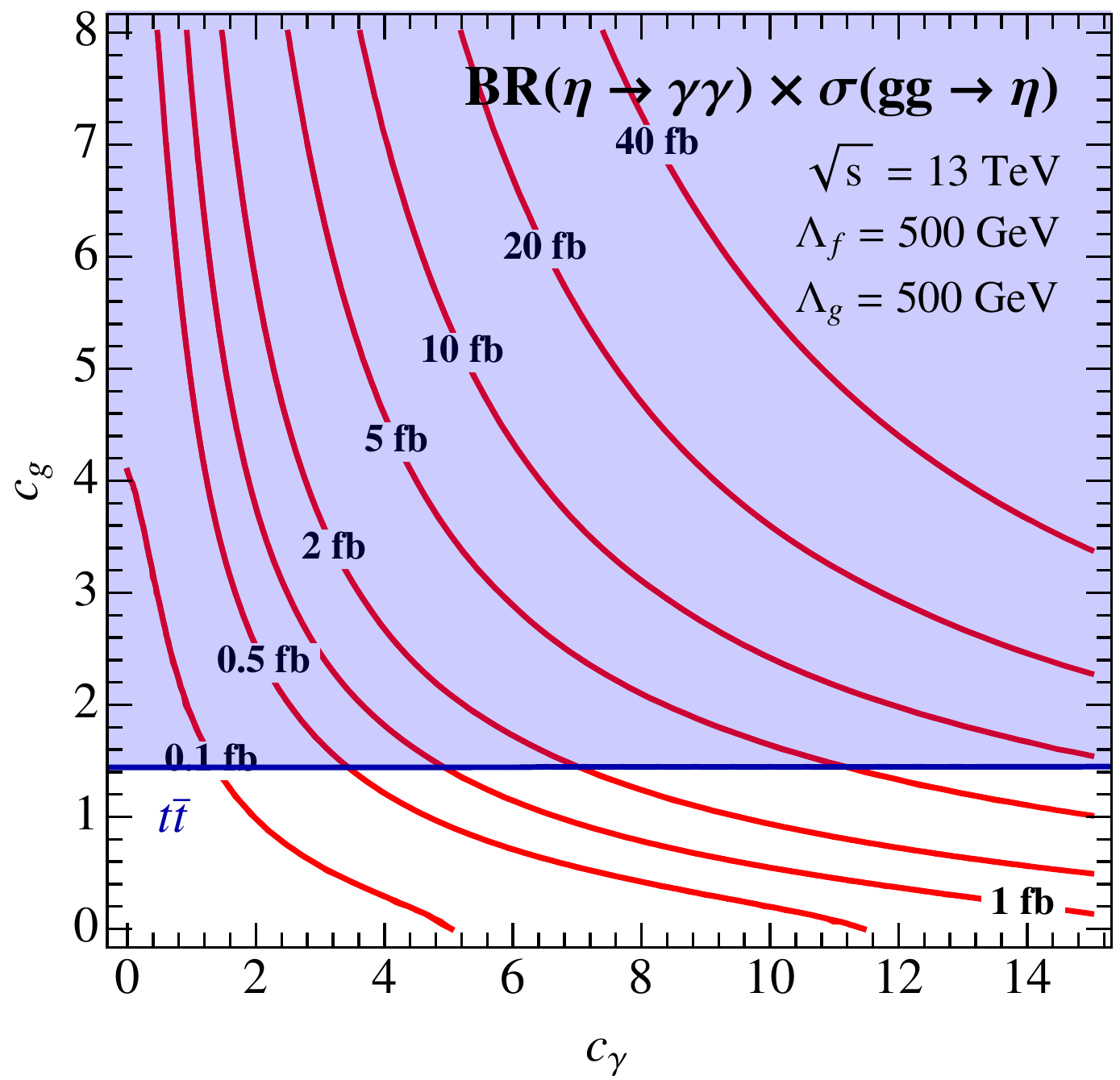}
\caption{The diphoton rate at 13 TeV using $\Lambda_f =$ 500 GeV and $\Lambda_g =$ 500 GeV.  The blue region is excluded by $t\bar{t}$ searches.}
\label{fig:diphotonrate1}
\end{center}
\end{figure}

The diphoton rate is computed in Fig.~\ref{fig:diphotonrate1} as a function of $c_\gamma$ and $c_g$.  Clearly $c_\gamma$ only affects the branching ratio, while $c_g$ both the total rate and the branching ratio since it modifies $\Gamma_{gg}$.  The blue shaded region indicates where the model is ruled out by 8 TeV searches for $t\bar{t}$ resonances (rescaled to 13 TeV).  One can see that a sufficient diphoton rate can be achieved by having either one of $c_\gamma$ or $c_g$ to be sizable, but because $c_g$ increases the total rate, the $t\bar{t}$ rate also increases.  Dijet searches also constrain $c_g < 6$ and $Z\gamma$ searches constrain the diphoton rate to be less than 20 fb.  These are not shown in Fig.~\ref{fig:diphotonrate1} since $t\bar{t}$ is stronger than both.

\begin{figure}[t]
\begin{center}
\includegraphics[width=.45\textwidth]{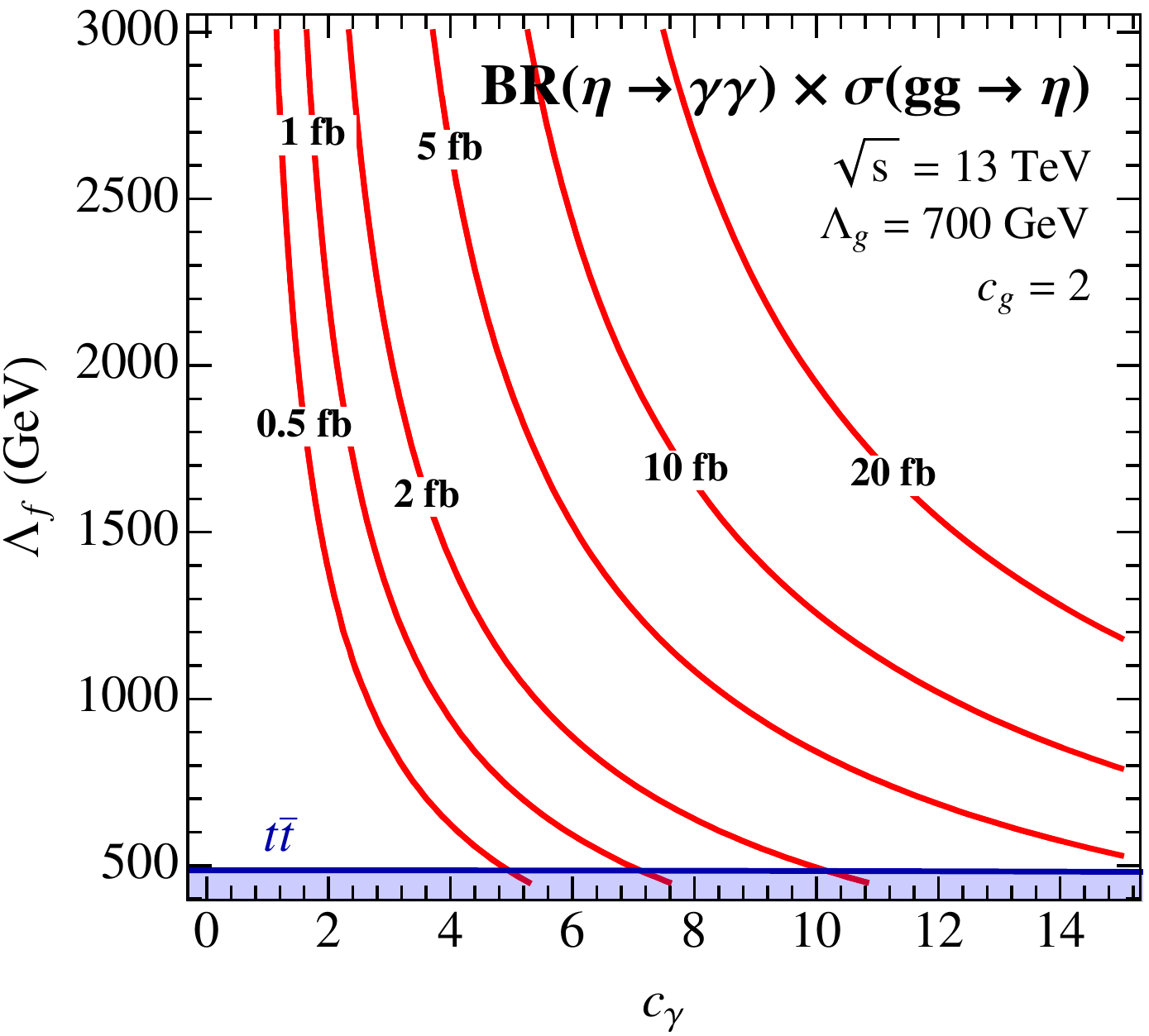}
\caption{The diphoton rate at 13 TeV using $\Lambda_g =$ 700 GeV and $c_g =$ 2.  The blue region is excluded by $t\bar{t}$ searches.}
\label{fig:diphotonrate2}
\end{center}
\end{figure}

In Fig.~\ref{fig:diphotonrate2} we slice the parameter space differently and fix a small contribution to gluon fusion via $c_g=2$ and look at the dependence on $\Lambda_f$.  We see that as $\Lambda_f$ is increased, the top loop contribution to the production shrinks as does the $t\bar{t}$ rate itself.  The appropriate rate is still attainable from the $c_g$ and $c_\gamma$ contributions.  Dijet searches are not constraining here because the overall rate is smaller and $Z\gamma$ searches still bound the overall diphoton rate (but is not shown in Fig.~\ref{fig:diphotonrate2}).

\subsection*{Results}

From Figs.~\ref{fig:diphotonrate1} and~\ref{fig:diphotonrate2} one can see that it is possible to achieve the observed signal rate of $\sim 1-5$ fb.  In both cases the strongest constraints come from $t\bar{t}$.  Dijet searches are not as sensitive nor are diboson searches as we have used safe value of $c_W=0$.  From the interplay of the effective operators of Eq.~\eqref{eq:interactions} two parameter regions that can explain the excess can be identified:

\begin{itemize}

\item {\it A single scale} where $\Lambda_f = \Lambda_g = f$ as in Fig.~\ref{fig:diphotonrate1}.  Given that the scales are not too large, the pseudoscalar to gluon coupling must come mainly from the top loop and one requires a large $c_\gamma$ value to get the diphoton rate.

\item {\it Suppressed fermions} where $\Lambda_f \gg \Lambda_g = f$ as in Fig.~\ref{fig:diphotonrate2}.  Here the pseudoscalar to top coupling is small enough that $t\bar{t}$ searches are not too constraining.  Then gluon fusion can receive a moderate enhancement and the pseudoscalar to photon coupling also only needs a moderate enhancement.

\end{itemize}

We use the scale $f$ to indicate the scale at which the dimension 5 operators are generated.  In the case of a suppressed fermion contribution one can achieve $\Lambda_f \gg \Lambda_g$ either by the fermion contribution being generated at a much higher scale or by a small prefactor such that $\Lambda_f \gg f$.  The latter case will be relevant for the composite Higgs case.

Another possibility for a large enough diphoton rate, not mentioned above, is to invoke a large contribution from $c_g$.  To avoid $t\bar{t}$ bounds, it is needed that $\Lambda_f \gg \Lambda_g$, making the coupling of $\eta$ to tops negligible for all practical purposes.  In this limit the rate no longer depends on $\Lambda_f$, but only on $(\Lambda_g, c_\gamma, c_g)$.  If, for illustration, one fixes $\Lambda_g=$ 700 GeV and the diphoton rate to 2 fb, then $c_g$ is the only free parameter since $c_\gamma$ is determined by the diphoton rate.  Then the dijet rate is $\approx c_g^2$~(400 fb) and the total width is $\approx c_g^2$~(0.04 GeV).  We see that in this case, dijet searches bound $c_g \lesssim 4$ and produce a narrow resonance.

\begin{figure}[t]
\begin{center}
\includegraphics[width=.45\textwidth]{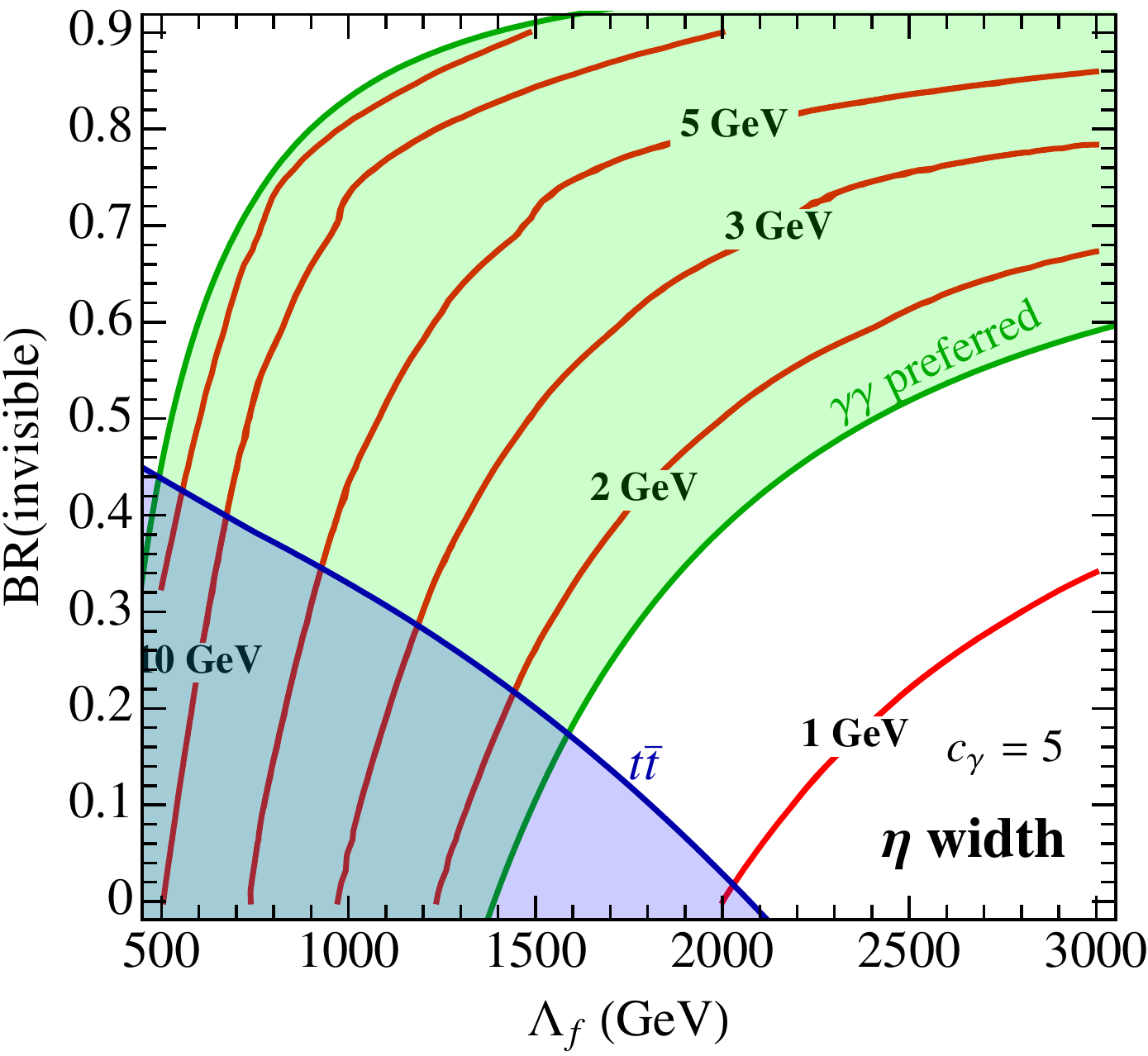} \quad\quad\quad
\includegraphics[width=.45\textwidth]{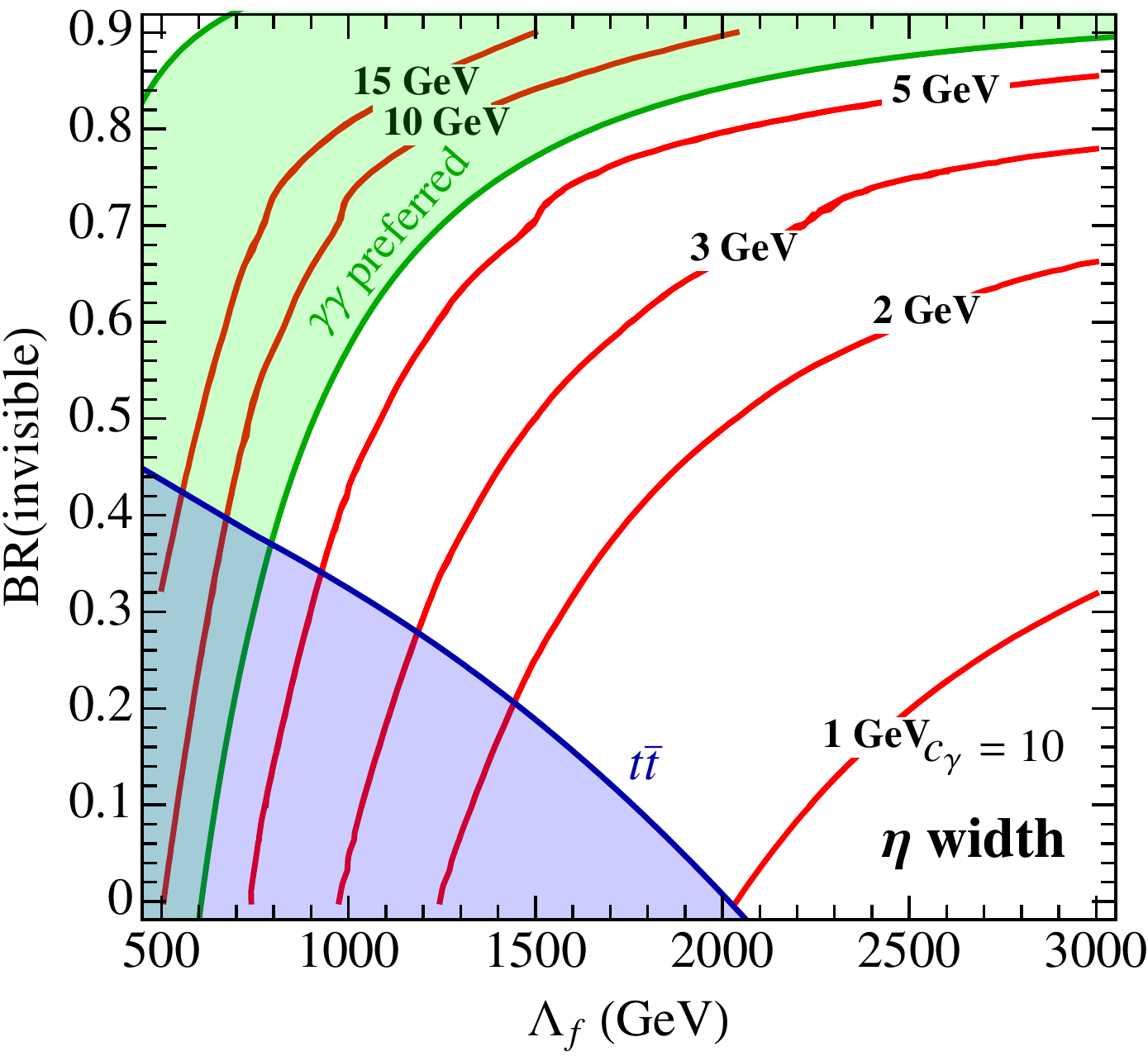}
\caption{The width of the pseudoscalar as a function of the fermion suppresion scale $\Lambda_f$ and the invisible branching ratio for $c_\gamma=5$ (left) and $c_\gamma=10$ (right).  The blue region is excluded by $t\bar{t}$ searches and the green region has a diphoton rate between 1 and 10 fb.  The parameters used are $\Lambda_g=$ 500 GeV and $c_g=2$.}
\label{fig:etawidth}
\end{center}
\end{figure}

\paragraph{}
Now that we have identified viable regions of parameter space we comment on the width in more detail than Eq.~\eqref{eq:width-estimate}.  Fig.~\ref{fig:etawidth} shows the width as a function of $\Lambda_f$ and the invisible branching ratio.  With only the SM states we have discussed, there is no invisible width and the $\eta$ tends to be narrow.  A wider resonance can be obtained by adding an invisible width.

\section{The mass scale of a pseudoscalar}\label{sec:pseudoscalarmass}

In this section we describe a model in which one can naturally find a pseudoscalar of mass $\sim$750 GeV.  In this model, both the Higgs and the $\eta$ are pNGBs of a global symmetry.  The argument is based on the composite Higgs scenario (for a nice review, see~\cite{Panico:2015jxa}) where the lightest particles of the composite sector are pNGBs.  The minimal case identifies the pNGB multiplet with the Higgs multiplet which crucially depends on the global symmetries~\cite{Agashe:2004rs}.  One can consider non-minimal scenarios, however, where there are additional light pNGBs which can have various quantum numbers and could even be SM singlets.  See~\cite{Gripaios:2009pe,Serra:2015xfa,Katz:2005au,Galloway:2010bp,Mrazek:2011iu} for previously studied examples.

\subsection*{The general framework}

In adding another light scalar, where light is relative to the cutoff, one is once again faced with a hierarchy problem.  Just as identifying the Higgs as a pNGB can explain its small mass, the presence of an additional $\sim$750 GeV pseudoscalar can be naturally justified if it is also a pNGB of a global symmetry.

In order to accommodate an extra singlet (or extra singlets) we need to go beyond the minimal composite Higgs model~\cite{Agashe:2004rs} and consider a larger global group $\mathcal{G}$.  The coset $\mathcal{G}/\mathcal{H}$ then contains the SM Higgs doublet and extra scalars.\footnote{A notable case is SO(6)/SO(5) with a Higgs and a pseudoscalar singlet~\cite{Gripaios:2009pe}, see App.~\ref{sec:app}.}  To control custodial breaking effects that may be induced by the additional scalars it is phenomenologically important to add extra discrete symmetries~\cite{Mrazek:2011iu}.

As we avoid discussion of a particular model, for our purposes it is sufficient to highlight a few generic facts for models with a pseudoscalar singlet pNGB in addition to the Higgs pNGB multiplet.  The full set of pNGBs can be parametrized as
\begin{equation}
U(\Pi)= \exp \left( \frac{i}{f} ( \hat{H} + \eta T_\eta + \ldots)\right),
\end{equation}
where $\hat{H}$ is a compact notation for the matrix of pNGBs that will be identified with the SM Higgs and $\eta$ is the pseudoscalar associated with the broken generator $T_\eta$.  The $\ldots$ indicate additional pNGBs that could be present.

The standard model SU(2)$_L$ $\times$ U(1)$_Y$ gauges a subgroup of the unbroken $\mathcal{H}$. In particular, for $\eta$ to be a singlet, we must have
\begin{equation} \label{eq:singlet}
\left[ T_\eta, T_{\mathrm{SM}} \right]=0,
\end{equation}
where $T_{\mathrm{SM}}$ are the generators corresponding to the SM gauge fields.  This has relevance for phenomenology, since as it is a singlet the $\eta$ does not couple to SM gauge fields.

The general couplings of the pNGBs to SM vectors are given by
\begin{equation} \label{eq:vectorcouplings}
\frac{g_{hVV}}{g_{hVV}^{\rm SM}}= 1 - \kappa_V\frac{v^2}{f^2} + \mathcal{O}\left(\frac{v^4}{f^4}\right),
\quad\quad\quad
\frac{g_{\eta VV}}{g_{hVV}^{\rm SM}}=0,
\end{equation}
where $g_{hVV}^{\rm SM}$ is the Higgs-vector-vector coupling in the standard model and $\kappa_V$ is an $\mathcal{O}(1)$ coefficient.  The pseudoscalar does not couple to SM vectors at tree level.  From Eq.~\eqref{eq:vectorcouplings} one can derive a lower bound on the scale $f$ which is found to be $f \gtrsim$ 600 GeV~\cite{ATLAS:2015bea,Khachatryan:2014jba}.  Another important implication of Eq.~\eqref{eq:vectorcouplings} is that the gauge interactions do not contribute to the one loop generation of a bare mass of the pseudoscalar.

\subsection*{The fermion sector}

At this point the couplings between the pNGBs and the SM fermions have not been specified.  In this work we focus primarily on the coupling of the pseudoscalar to the top quark because it has the largest Yukawa coupling.  The usual generation of masses for SM quarks in composite Higgs models proceeds via the partial compositeness mechanism~\cite{Kaplan:1991dc} where the elementary fields couple to operators from the composite sector.  Schematically the coupling is
\begin{equation}\label{eq:pc}
y_L \overline{q_L}\cdot U \cdot\Psi
+ y_R \overline{u_R}\cdot U \cdot \Psi + h.c.,
\end{equation}  
where $\Psi$ represent composite operators and $y_L$ and $y_R$ are related to the fermion Yukawas.  While Eq.~\eqref{eq:pc} can be made formally non linearly invariant under $\mathcal{G}$, the SM fermions are embedded in incomplete multiplets of $\mathcal{G}$ which breaks the global symmetries.  This breaking in turn generates Yukawa couplings and a potential for the pNGBs.  Generically, the Higgs potential always receives a contribution from at least the left handed mixing.

The interactions of the singlet, on the other hand, are model dependent.  In particular, if the embeddings of $q_L$ and/or $u_R$ are not eigenstates of the generator $T_\eta$, then in general the interactions of Eq.~\eqref{eq:pc} break the shift symmetry of $\eta$ and contribute to its potential.  It is also important to ensure that the embeddings are consistent with our assumption of CP conservation.  It has been shown that this can be done in concrete examples~\cite{Gripaios:2009pe}.

By the appropriate insertions of spurions, $y_L$ and $y_R$, we can construct the would-be Yukawa term
\begin{equation} \label{eq:topmass-ch}
y_t \overline{t_L} h t_R \left( 1 + i \kappa_\eta \frac{\eta}{f}  + {\mathcal{O}}\left( \frac{1}{f^2}\right) \right) + h.c.,
\end{equation}
where $\kappa_\eta$ is an $\mathcal{O}(1)$ coefficient.

The couplings of the $h$ and $\eta$ to top quarks is found to be
\begin{equation}\label{eq:fermion-couplings}
\frac{g_{htt}}{g_{htt}^{\rm SM}} = 1 - \kappa_F \frac{v^2}{f^2},
\quad\quad\quad
\frac{g_{\eta tt}}{g_{htt}^{\rm SM}}=i \frac{v}{f}\kappa_\eta.
\end{equation}
where $g_{htt}^{\rm SM}$ is the top coupling to the Higgs in the standard model and $\kappa_F$ is an $\mathcal{O}(1)$ coefficient that depends on the embedding of the fermions.  Notice that derivation has been completely general, and the only assumptions have been related to the CP nature of the singlet. It is also manifest that, from the SM perspective, the coupling of the $\eta$ arises at dimension 5 in complete analogy with the simplified discussion of Sec.~\ref{sec:interactions}.

\subsection*{Mass of the pseudoscalar}

The mass of the $\eta$ is determined by the parameter that breaks its shift symmetry.  Even though the $\eta$ is an SM singlet, if the embeddings of $q_L$ or $u_R$ break $T_\eta$, then the $\eta$'s shift symmetry will be broken.  Then Eq.~\eqref{eq:topmass-ch} will contribute to the $\eta$'s mass via a contribution to $\lambda_{\eta h}$.  This contribution is chirality breaking and involves a Higgs field.  There is a chirality preserving contribution that we expect to directly contribute to $m_\eta^2$ and arises in the following way.

After having integrated out the composite sector at low energies for $u_R$ we have
\begin{equation} \label{eq:formfactor}
\overline{u_R} \;\slashed{p}\; u_R + y_R^2 F_{u_R}(p^2,m_*) \overline{u_R} \;\slashed{p}\; u_R \left(c_\eta \frac{\eta^2}{f^2} + \ldots \right),
\end{equation}
where $F_{u_R}$ is a form factor that encodes the contribution of the resonances of the strong sector.  The poles of $F_{u_R}$ correspond to the masses of the resonances of the strong sector.  Here we use $m_*$ to denote the various mass scales of the resonances that we expect below $4\pi f$, but above $f$.

Note that Eq.~\eqref{eq:formfactor} is generic for pNGBs that couple to $u_R$.  It is possible that in specific models $c_\eta$ can vanish due to accidental symmetries~\cite{Gripaios:2009pe,Mrazek:2011iu}.  In other models $c_\eta$ can be proportional to $\kappa_\eta$.  Here we simply consider it to be an $\mathcal{O}(1)$ coefficient.  We find a term in the effective potential of the form
\begin{equation} \label{eq:mass-term-pot}
c_\eta \frac{N_c y_R^2}{4\pi^2}m_*^2 \eta^2.
\end{equation}
Fixing the top Yukawa, we find
\begin{equation}
y_t \simeq \frac{f}{m_*} y_L y_R
\end{equation}
and taking $y_L \sim y_R$ we arrive at the estimate,
\begin{equation}
m_\eta^2 \simeq \frac{N_c y_t}{2\pi^2} \frac{m_*^3}{f}.
\end{equation}
For reasonable values of the parameters we get the estimate,
\begin{equation}\label{eq:mass-eta}
m_\eta \simeq 750~\mathrm{GeV} \left(\frac{m_*}{1.3~\rm TeV}\right)^{3/2} \left(\frac{600~\mathrm{GeV}}{f}\right)^{1/2}.
\end{equation}
Interestingly, this is of the right size.  It is worth further emphasizing that this mass is at the naturalness limit for $\eta$ since no tuning is required.  This is different than for the Higgs where we need to tune a different combination of coefficients involving the left handed quarks and gauge fields to successfully achieve electroweak symmetry breaking, $v \ll f$.

In this respect, we usually expect a ratio given by
\begin{equation}\label{eq:ratio-masses}
\frac{m_\eta}{m_h} \sim \sqrt{\frac{g_*}{y_t}}\frac{f}{v}. 
\end{equation}
Notice that the usual tuning in composite Higgs models requires $g_* \simeq m_*/f \simeq \mathcal{O}(1)$, {\it i.e.} top partners within reach of the LHC.  The same prediction derived from the Higgs mass is true in this model from the $\eta$ mass. Models of this type, where the mass of the new resonance is natural and linked to the explanation of size of the Higgs mass, seem to deserve further attention even if one has to introduce new ingredients on top of the minimal models.

\subsection*{Interactions of the singlet}

In order to connect the composite $\eta$ with the results of Sec.~\ref{sec:interactions} we comment on the size of $c_\gamma / \Lambda_g$.  We start with the top coupling, which from Eq.~\eqref{eq:fermion-couplings}, tells us that tops will couple to the pseudoscalar with a $v/f$ suppression according to
\begin{equation}\label{mathching-f}
\frac{1}{\Lambda_f} \simeq \frac{\kappa_\eta}{f}.
\end{equation}
In the limit where the $\eta$ is the lightest new state, the loop induced couplings to gluons and photons are dominated by top contributions.

In the composite sector there are particles (the top partners) charged under both SU(3) and electromagnetism that can also run in the loop.  From the view of the composite sector, $\eta$ is a NGB which means that any shift breaking interaction with top partners must go through an elementary composite mixing.  For an estimate, we note that each power of $\eta$ comes with at least one power of $y \sim y_R \sim y_L$.  Given the symmetries of the strong sector, some of these corrections can have further suppressions.  For estimates see App.~\ref{sec:app}.

The challenge of finding large enough $c_\gamma$ presents itself from the fact that top partner searches have been performed and it seems difficult to evade a bound of $\sim$700 GeV (see for example~\cite{Matsedonskyi:2014mna,Aguilar-Saavedra:2013qpa}) and go into a region where the loop functions are enhanced.  A similar scaling is expected for the top partner contribution to $c_g$ (without the color and electric charge factors).

It is possible that the global (non-linearly realized) symmetry of the strong group is anomalous.  In the case where the generator associated to $\eta$ has non-vanishing anomaly coefficients with two SM gauge bosons, one can have dimension 5 operators in complete analogy with Eq.~\eqref{eq:interactions}.  The simplest scenario with a light singlet, SO(6) $\simeq$ SU(4), can have global anomalies (SU(4)$^3$) although with $c_\gamma=0$ (see App.~\ref{sec:app}).

Other coset spaces can have additional singlets, an example (that suffers custodial breaking and hence is tuned) is SU(3)$\times$U(1)$_X$ / SU(2)$_L$$\times$U(1)$_Y$~\cite{Contino:2003ve,Schmaltz:2004de}.  The NGBs are in the $\mathbf{2}_{\pm 1/2} + \mathbf{1}_0$ of SU(2)$_L$$\times$U(1)$_Y$.  In this case $T_\eta\sim \mathrm{diag}(0,0,1)$.  The presence of the U(1)$_X$ allows for a correct hypercharge assignment and the NGBs have charge $X=-1/3$. Hypercharge is defined as $Y=(1/2\sqrt{3})\lambda_8 + X$, where $\lambda_a$ are the Gell-Mann matrices. In this case $T_{\rm em}^2$ has no particular structure and $\mathrm{tr}[T_\eta T_{\rm em}^2]\neq 0$ in general. At low energies this can manifest itself into an anomalous contribution in the form
\begin{equation}
n_\gamma \frac{\alpha}{4\pi} \frac{\eta}{f} F_{\mu\nu} \tilde{F}^{\mu\nu},
\end{equation}
which can help numerically to get a sizable decay to photon pairs.  Other choices of global groups could give the same contribution (finding these groups could be a direction of further study), and more exotic groups can also contain color anomalies, hence a contribution to $c_g$.  Finally, notice that in this case the anomaly coefficient is not suppressed by SM couplings.

\subsection*{Results}

As a brief summary of the possibilities discussed, we comment on two specific cases, both using a moderate scale $f \sim$ 600 GeV, as is suggested by Higgs coupling measurements and naturalness considerations.  In particular, given the notation of Eq.~\eqref{eq:interactions} we consider $\Lambda_g=f$ and $\Lambda_f=f/\kappa_\eta$ where $\kappa_\eta$ is defined in Eq.~\eqref{eq:fermion-couplings}. We leave $c_\gamma$ and $c_g$ as free parameters, having in mind the possible size as suggested by the previous estimates.

\begin{itemize}

\item $\kappa_\eta$ is $\mathcal{O}(1)$.  In this case where $\Lambda_f=\Lambda_g= f$ in order to sufficiently enhance the diphoton rate a large anomalous contribution to the diphoton coupling seems necessary.

\item $\kappa_\eta$ is reduced.  Then $\Lambda_f > \Lambda_g = f$ and we can be in the case where $\Lambda_f \simeq$ 3 TeV where only moderate values of $c_\gamma$ and $c_g$ are required (see Fig.~\ref{fig:diphotonrate2}).  This is probably still difficult to achieve in the minimal realizations of composite Higgs which only include tops and top partners.

\end{itemize}

\begin{figure}[t]
\begin{center}
\includegraphics[width=.5\textwidth]{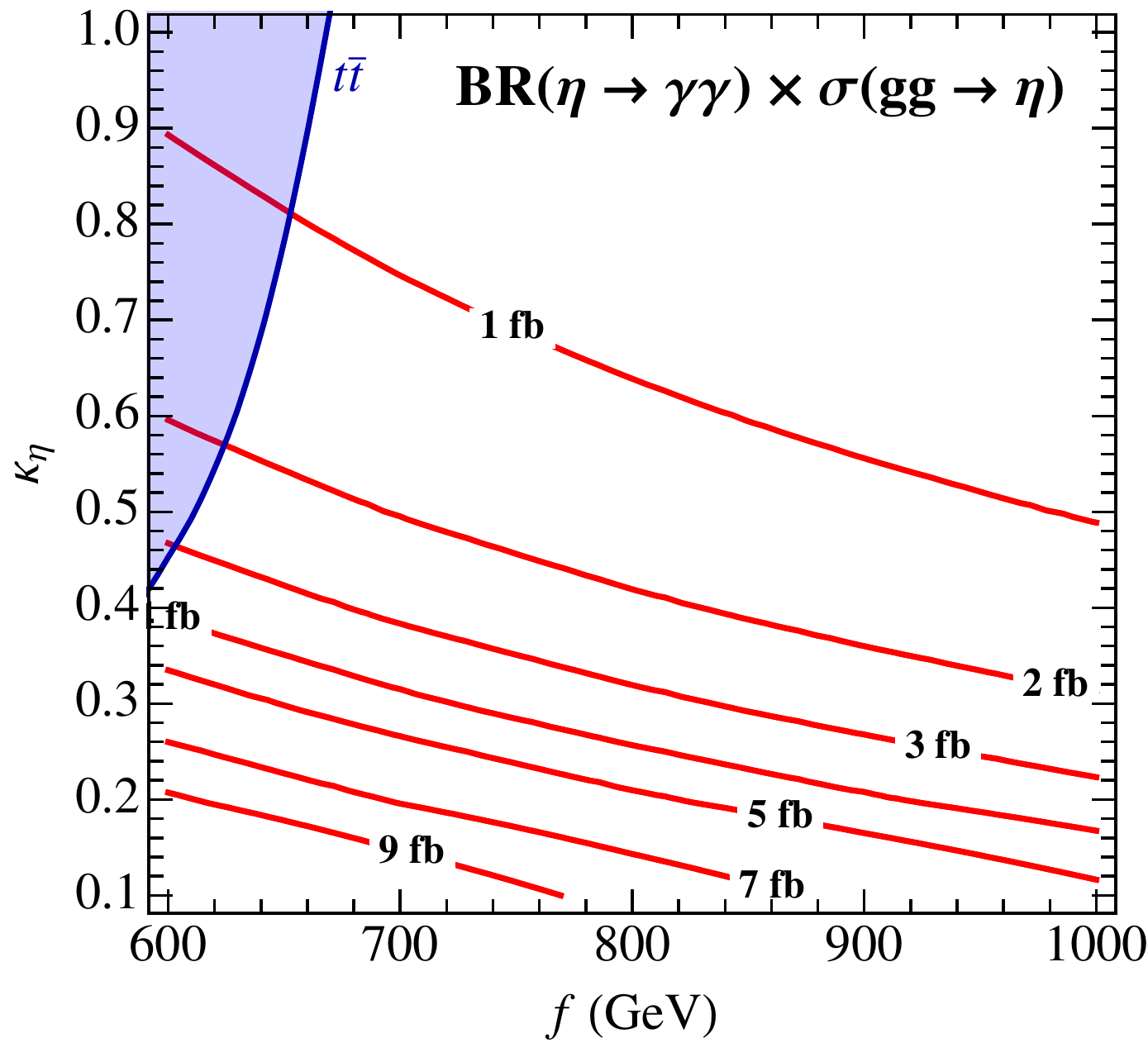}
\caption{The diphoton rate at 13 TeV for the composite Higgs scenario.  The parameters $c_g = 2$ and $c_\gamma = 2$ are used.}
\label{fig:composite}
\end{center}
\end{figure}

Both of these cases can be visualized in Fig.~\ref{fig:composite}, where we have fixed $c_\gamma=2$ and $c_g=2$.  Near the top at $\kappa_\eta \simeq 1$ the rate is too low with $c_\gamma$ and $t\bar{t}$ forces $f$ to start to become large.  For small $\kappa_\eta$ both the diphoton rate is sufficient and $f$ can be near the preferred value.

\section{Alternative explanations}\label{sec:alternatives}

While we have focused on the case of a pseudoscalar resonance, there are obviously a number of possible explanations.  The general obstacles that models face were also shared in the pseudoscalar case which are boosting the $\gamma\gamma$ branching ratio to $\sim 10^{-2}$ and avoiding constraints from $WW$, $ZZ$, $Z\gamma$, and $hh$ searches.  In fact, a nice feature of the pseudoscalar explanation is that symmetries enforce a loop level coupling to both photons and transverse vector bosons, easily evading diboson constraints.

In this section, we briefly outline two other scenarios that could be plausible, namely a spin-0 scalar resonance and a spin-2 resonance.  There are other scenarios one can envisage which we do not comment on at all.  One example of this would be a 750 GeV particle decaying to two $\mathcal{O}(100~\rm MeV)$ particles that each decay to photon pairs.  The large boost of the light particles then cause the pair of photons to be detected as a single photon.

\subsection{Scalar resonance}

A model very similar to the pseudoscalar is a scalar singlet $s$ added to the SM.  While assuming CP allowed us to restrict the pseudoscalar from mixing with the SM, we do not have such a symmetry for the scalar (since $Z_2$ is not useful in this context).  In any case, it is possible to assume that the only couplings of the scalar $s$ to the SM are through $F_{\mu\nu} F^{\mu\nu}$ and $G_{\mu\nu}^a G^{\mu\nu a}$, possibly induced by heavy vector-like fermions (see {\it e.g.}~\cite{Aguilar-Saavedra:2013qpa}).  As discussed in Sec.~\ref{sec:interactions} we find it useful to work below the scale of the new fermions (in order to avoid decays to them).\footnote{As the heavy fermions must be colored to couple to gluons, they also must be unstable because they are colored.  Additional model building is necessary to ensure these are phenomenologically viable and will likely lead to specific predictions of the mass or couplings.}  This model is parametrized by only two interactions (for simplicity, we neglect here constraints from $Z\gamma$, see Sec.~\ref{sec:lhc})
\begin{equation}
\mathcal{L} = \frac{\alpha}{4\pi}\frac{s}{\Lambda_F} F_{\mu\nu} F^{\mu\nu} + \frac{\alpha_s}{4\pi} \frac{s}{\Lambda_G} G_{\mu\nu}^a G^{\mu\nu a}.
\end{equation}
The overall rate is a function of $\Lambda_G$ and can be estimated by rescaling the SM rate (similar to Sec.~\ref{sec:interactions})
\begin{equation}
\sigma_s(750~\mathrm{GeV}) = \left(\frac{v}{\Lambda_G}\right)^2 \times 1.25~\mathrm{pb}.
\end{equation}
The branching ratios are functions of both scales (and gauge couplings) and are
\begin{equation}\begin{aligned}
\br_{s\to gg}           &= \frac{8\alpha_s^2}{8\alpha_s^2 + \alpha^2 (\nicefrac{\Lambda_G}{\Lambda_F})^2}, \\
\br_{s\to \gamma\gamma} &= \frac{\alpha^2 (\nicefrac{\Lambda_G}{\Lambda_F})^2}{8\alpha_s^2 + \alpha^2 (\nicefrac{\Lambda_G}{\Lambda_F})^2},
\end{aligned}\end{equation}
which can give $\br_{s \to \gamma\gamma} \sim 10^{-2}$ when $\Lambda_G$ is sufficiently larger than $\Lambda_F$.  Then one needs to adjust the cross section with $\Lambda_G$ where the right value is $\Lambda_G/v \simeq 5$.  A full analysis would involve a study of the heavy fermions, including their impact on the running of gauge and Yukawa couplings (which is especially important if the two loop induced effects on the Higgs quartic are large).  Such studies are model dependent but would allow one to make concrete predictions for accompanying signals.

\subsection{Spin-2 resonance}

There are no obstructions for spin-2 particles to decay to photon pairs.  Here we consider a hypothetical massive spin-2 particle $\rho_{\mu\nu}$ with a mass $m_\rho$ that couples to the SM stress energy tensor.  While it is debatable whether such a light spin-2 particle could be the first observed state ({\it e.g.} this is not the case in QCD), we explore this possibility with a very pragmatic approach.

Consider a spin-2 Lagrangian with a Fierz-Pauli mass term (as is automatically implied by the Kaluza Klein reduction of 5 dimensional models, see~\cite{Hinterbichler:2011tt} for a review)
\begin{equation}
\mathcal{L} = \mathcal{L}_{\rm spin-2}^{\rm FP} + \frac{1}{\Lambda_G} \rho_{\mu\nu} T^{\mu\nu}_B + \frac{1}{\Lambda_F} \rho_{\mu\nu} T^{\mu\nu}_F,
\end{equation}
where $\rho_{\mu\nu}$ is canonically normalized.  We have also separated the interactions with the gauge fields and the Higgs $T^{\mu\nu}_B$ from the stress energy tensor of fermions $T^{\mu\nu}_F$.   Other separations could be possible, but for simplicity we use this distinction.

In the limit $m_\rho \gg m_{\mathrm{SM}}$ the form of the partial widths are dominated practically by counting degrees of freedom, since the leading contributions from mass terms are proportional to $T^{\mu\nu} \sim \eta^{\mu\nu}$ and thus vanish onshell.  The partial decays widths are~\cite{Han:1998sg}
\begin{equation}
\Gamma_{\gamma\gamma} \simeq \frac{m_\rho^3}{80\pi \Lambda_G^2},
\quad\quad
\Gamma_{gg} \simeq 8 \Gamma_{\gamma\gamma},
\quad\quad
\Gamma_{ZZ} \simeq \frac{13}{12} \Gamma_{\gamma\gamma},
\quad\quad
\Gamma_{WW} \simeq \frac{13}{6} \Gamma_{\gamma\gamma},
\quad\quad
\Gamma_{hh} \simeq \frac{1}{12} \Gamma_{\gamma\gamma},
\end{equation}
and
\begin{equation}
\Gamma_{\ell\ell} \simeq \frac{m_\rho^3}{160\pi \Lambda_F^2},
\quad\quad
\Gamma_{qq} \simeq N_c \Gamma_{\ell\ell}.
\end{equation}
If the ratio of couplings is $\Lambda_G / \Lambda_F \ll 1$ the dominant production channel could be gluon fusion.  Given that the ratios among boson couplings have been fixed, the diphoton branching ratio is
\begin{equation}
\br_{\rho \to \gamma\gamma} \simeq \frac{3}{37} + \mathcal{O}\left(\frac{\Lambda_G^2}{\Lambda_F^2}\right)
\sim 8\%.
\end{equation}
The total rate should thus be $20-40$ fb.  Due to the sensitivity of dilepton searches, the branching ratios of leptons must be $\lesssim 1\%$ which justifies the approximation made.  The total rate for fixed $m_\rho$ then is a function of $\Lambda_G$ which can be selected to achieve the correct rate to explain the excess.

\section{Conclusions}\label{sec:conclusions}

In this paper we considered a possible framework that can explain the excess reported by ATLAS and CMS in the search for diphoton resonances and and explored the consequences.  Given the challenges imposed by a resonance observed first in its decay to diphoton, a channel with a notoriously small branching ratio, we focused on a pseudoscalar resonance.  This is a scenario where we avoid the very strong limits posed by dilepton and diboson searches (see Table~\ref{tab:limits13tev}).  As discussed, a proposed singlet pseudoscalar couples to SM particles only at the non-renormalizable level to operators that are SM singlets and CP odd.  This list is rather constrained and the leading interactions are to tops, gluons, and photons.  We emphasized two possible limits of the effective description.  Particles of other natures, like a CP even scalar or a spin-2 particle also face numerous constraints.

The first limit, a ``natural scenario'', could be offered by new physics in which the suppression scale is common for all operators, thus establishing the coupling of the singlet to the SM top as the leading interaction.  In this limit we showed that searches for $t\bar{t}$ resonances from 8 TeV data already constrain part of the parameter space.  Moreover, in this case to match the diphoton rate, the coupling to photons requires a sizable contribution from new physics.  This case seems difficult to realize in composite Higgs models without adding new states in addition to those from the composite sector.
As we have stressed, the challenge is to get a sufficiently large branching ratio to diphoton which could be done by relying on anomalous couplings which are allowed if the global symmetries of the composite sector are anomalous.

The second limit is to assume that the new physics responsible for the effective operators only produces a sizable coupling to the field strengths of the gauge bosons (and not to the fermions).  In this case the $t\bar{t}$ constraint is avoided and the excess can be reproduced by invoking sizable effects in $G_{\mu\nu}^a \tilde{G}^{\mu\nu a}$.  Then the suppression scale of $\sim$3 TeV for the fermion operator (and $\mathcal{O}(1)$ coupling) could be sufficient to explain the excess.  Since only moderate values of $c_\gamma$ and $c_g$ are required, fewer new states are needed relative to the previous case.  One drawback is that even if this scenario was realized in the composite picture, one still requires a moderate tuning of the size of the coupling between the singlet and the top.

A common aspect to both viable scenarios is that the diphoton excess can only be explained if a sector of new particles and interactions is present at a relative low scale, comparable or possibly even lower than 750 GeV.  In particular the presence of colored and electroweakly charged states seems unavoidable. Moreover, in the case of composite models with anomalous contributions to the decay channels, one expects colored (and possibly long lived) pions that might be accessible at LHC.  The diphoton excess represents an exciting prospect as Run 2 has only just started.  Forthcoming data will tell us more.

\subsubsection*{Acknowledgements}

The authors would like to thank Nima Arkani-Hamed, Gustavo Burdman, Raffaele Tito D'Agnolo, and David Pinner for useful conversations.  AT thanks Marco Farina and Michele Redi for interesting comments. ML is supported by a Peter Svennilson Membership at the Institute for Advanced Study, AT is supported by an Oehme Fellowship, and LTW is supported by DOE grant DE-SC0013642.

\paragraph*{The Arxiv excess} 
The diphoton excess has already received much attention~\cite{Harigaya:2015ezk,Mambrini:2015wyu,Backovic:2015fnp,Angelescu:2015uiz,Nakai:2015ptz,Knapen:2015dap,Buttazzo:2015txu,Pilaftsis:2015ycr,Franceschini:2015kwy,DiChiara:2015vdm}.  For issues related to a possible interpretation in composite Higgs scenarios \cite{Franceschini:2015kwy}.

\appendix
\section{SO(6)/SO(5) model}\label{sec:app}

In this appendix we derive some useful formulas for the SO(6)/SO(5) composite Higgs model with a Higgs and a pseudoscalar $\eta$.  This section is intended to clarify some of the estimates and arguments given in Sec.~\ref{sec:pseudoscalarmass} with the aid of an explicit case.

\subsection{The gauge sector}

The standard model SU(2)$_L$ $\times$ U(1)$_Y$ is embedded in an SO(4) subgroup of the unbroken SO(5), under which the $\eta$ is an exact goldstone. The ``pions'' of the coset space can be organized in the matrix
\begin{equation}
\Pi = \sqrt{2}(h^i T^i + \eta T_\eta)
= \left(\begin{array}{cc|c} \mathbf{0}_4 & 0 & \vec{h} \\ 0 & 0 & \eta \\ \hline - \vec{h}^T & -\eta & 0
\end{array}\right),
\end{equation}
where $T^i$ ($i=1,2,3,4$) and $T_\eta$ are the broken generators of SO(6).  For convention we gauge the SU(2)$_L$ $\times$ U(1)$_Y$ subgroup in the upper left $4 \times 4$ block, which is consistent with the assumption of $\eta$ being a SM singlet.  We then define the vector $\Sigma_i \equiv {[\exp(i\Pi/f)]_i}^6$ as (where now $i=1,\ldots,6$)
\begin{equation}
\Sigma^T=  \frac{\sin(\nicefrac{\sqrt{h^2+\eta^2}}{f})}{\nicefrac{\sqrt{h^2+\eta^2}}{f}}
(\vec{h},\eta, f \cot(\nicefrac{\sqrt{h^2+\eta^2}}{f})),
\end{equation}
where $h^2 = \vec{h}^2$.  We perform a field redefinition~\cite{Gripaios:2009pe}
\begin{equation} \label{eqA:shift}
\vec{h} \leftarrow f\frac{\vec{h}}{\sqrt{h^2+\eta^2}}s,
\quad\quad\quad
\eta \leftarrow f\frac{\eta}{\sqrt{h^2+\eta^2}}s,
\end{equation}
where $s = \sin(\nicefrac{\sqrt{h^2+\eta^2}}{f})$.  In terms of the new fields, the goldstone multiplet is
\begin{equation}
\Sigma^T = (\vec{h}, \eta, \sqrt{f^2 - h^2 -\eta^2}),
\end{equation}
which leads to the following effective lagrangian (in unitary gauge)
\begin{equation} \label{eqA:beforeEWSB}\begin{aligned}
\mathcal{L} &= \frac{1}{2}(D^\mu \Sigma)^T D_\mu \Sigma \\
            &= \frac{1}{2}(\partial_\mu h)^2 + \frac{1}{2}(\partial_\mu \eta)^2 + \frac{1}{2f^2}(h \partial_\mu h +\eta \partial_\mu \eta)^2 + \frac{g^2}{4}h^2 W_\mu^a W^{\mu a} + \ldots
\end{aligned}\end{equation}
In this basis it is manifest that $\langle h \rangle = v =$ 246 GeV and the $\eta$ does not contribute to the electroweak vacuum expectation value.  From Eq.~\eqref{eqA:beforeEWSB}, however, we see that after electroweak symmetry breaking there will be a non canonically normalized kinetic term for $h$.  The following shift
\begin{equation} \label{eqA:canonicalnormalization}
h \to v + \sqrt{1-\frac{v^2}{f^2}}~h, 
\quad\quad\quad
\eta \to \eta, 
\end{equation}
restores canonical normalization and allows us to compute the couplings to vectors
\begin{equation} \label{eqA:vectorcouplings}
\frac{g_{hVV}}{g_{hVV}^{\rm SM}}= \sqrt{1-\frac{v^2}{f^2}},
\quad\quad\quad
\frac{g_{\eta VV}}{g_{hVV}^{\rm SM}}=0,
\end{equation}
where $g_{hVV}^{\rm SM}$ is the SM coupling.

\subsubsection*{Anomalies}

The global symmetry SO(6) $\simeq$ SU(4) can have anomalies. In terms of the SU(4) generators the embedding of the SM is 
\begin{equation}
T^a_L \sim \left(\begin{array}{cc} \sigma^a & \\ & \end{array}\right), 
\quad\quad\quad
T^a_R \sim \left(\begin{array}{cc} & \\ & \sigma^a\end{array}\right),
\quad\quad\quad
T_\eta \sim \left(\begin{array}{cc} \mathbb{I} & \\ & -\mathbb{I}\end{array}\right),
\end{equation}
while U(1)$_X$ is an external abelian factor.  Global anomalies of SU(4)$^3$ induce anomalous couplings of the $\eta$ to SM gauge fields, with the anomaly coefficients of SU(2)$_L$, $c_W$, and hypercharge, $c_B$, fixed by the embedding of the SM inside SU(4) to satisfy $c_W + c_B =0$, as can be explicitly checked. Indeed, the generator of the singlet is $T_\eta\sim \mathrm{diag}(1,1,-1,-1)$ while $T_{\rm em}\sim \mathrm{diag}(1,-1,1,-1) + q_X\mathbb{I}$, where $q_X$ is a charge of an additional $U(1)_X$.

\subsection{The fermion sector}

As discussed in Eq.~\eqref{eq:pc} the SM fermions are embedded in incomplete representations of SO(6).  More precisely the global group needs to be SO(6) $\times$ U(1)$_X$ where SM hypercharge is defined as $Y = X + T^3_R$. Among the several irreducible representations of SO(6), we consider here the \textbf{6}$_{2/3}$ which decomposes under SO(5) $\times$ U(1)$_X$ as a \textbf{5}$_{2/3}$ + \textbf{1}$_{2/3}$ + \textbf{1}$_{2/3}$.  Under the SU(2)$_L$ $\times$ SU(2)$_R$ the decomposition is
\begin{equation}
\mathbf{6}_{2/3} \to (\mathbf{2}, \mathbf{2})_{2/3} + (\mathbf{1}, \mathbf{1})_{2/3} + (\mathbf{1}, \mathbf{1})_{2/3} .
\end{equation}
We can embed quark doublet $q_L$ in the bidoublet component, while the $u_R$ can be embedded in a linear combination of the two singlets.  An embedding that is consistent with our assumption of CP conservation, also at the level of the composite sector, is %
\begin{equation} \label{eqA:quarkembedding}
q_L^T = \frac{1}{\sqrt{2}} (i b_L, b_L, it_L, -t_L,0,0) ,
\quad\quad\quad
u_R^T = (0,0,0,0,i\cos\theta,\sin\theta)~u_R .
\end{equation}
Eq.~\eqref{eqA:quarkembedding} shows that the mixing of $q_L$ does not break the shift symmetry of $\eta$ ({\it i.e.} $T_\eta q_L =0$) while in general the mixing of $u_R$ does break it.  Depending on value of $\theta$, which controls the coupling of $u_R$ to $\eta$, one can have different scenarios. 

For $\theta = \pi/4$, $\eta$ is an exact goldstone since the mixing respects the U(1)$_\eta$ symmetry that is generated by $T_\eta$.  Even though it is a goldstone it still couples to $u_R$.  On the other hand, for $\theta = \pi/2$, the mixing respects a discrete $Z_2$ symmetry, but $\eta$ does not couple to fermions. In the discussion in Sec.~\ref{sec:pseudoscalarmass} we implicitly avoided these two limiting cases to ensure a coupling between the $\eta$ and $t\bar{t}$. Lighter quarks, however, can have different embeddings and one can even choose embeddings to automatically satsify $\theta = \pi/2$ or $\theta =\pi/4$~\cite{Frigerio:2012uc}.

The form of the Yukawa term is constrained to be
\begin{equation}
y_t (q_L \Sigma) (\Sigma^T u_R)
= \frac{y_t}{\sqrt{2}} \overline{t_L} h t_R \left( \frac{\sqrt{f^2-h^2-\eta^2}}{f^2} \sin\theta + \frac{\eta}{f} i \cos\theta \right) + h.c.
\end{equation}
Normalizing to the SM Higgs couplings we have the following couplings to fermions for $h$ and $\eta$
\begin{equation}\label{eqA:fermion-couplings}
\frac{g_{htt}}{g_{htt}^{\rm SM}}= \frac{1-\nicefrac{2v^2}{f^2}}{\sqrt{1-\nicefrac{v^2}{f^2}}},
\quad\quad\quad
\frac{g_{\eta tt}}{g_{htt}^{\rm SM}}=i \frac{v}{f}\frac{\cot\theta}{\sqrt{1-\nicefrac{v^2}{f^2}}},
\end{equation}
to be compared with Eq.~\eqref{eq:fermion-couplings}.  Notice that since the top mass is proportional to $\sin\theta$ smaller values of $\sin\theta$ will increase the coupling of the top to the pseudoscalar, but will also induce tuning among the parameters of the model. 

Other terms can be written with the $\Sigma$, an example used in Sec.~\ref{sec:pseudoscalarmass} are chirality preserving operators that can induce a leading contribution to the potential for $\eta$, as
\begin{equation}\label{eqA:newkinetic}
\overline{u_R}\;\slashed{p}\;\Sigma^T \Sigma u_R 
= \overline{u_R}\;\slashed{p}\;u_R (\eta^2 \cos^2\theta  + (f^2-h^2-\eta^2) \sin^2\theta),
\end{equation}
which justifies the expression in Eq.~\eqref{eq:mass-term-pot}.

\subsection*{Contributions to $c_\gamma$ and $c_g$ from top partners}

A refined estimate for the UV contribution to $c_\gamma$ and $c_g$ from the top partners involves the full mass spectrum of the heavy fermions.  In order to be explicit, we consider the case where the left handed and right handed elementary quarks each couple to a \textbf{6} of SO(6).  The \textbf{6} decomposes as a \textbf{5} + \textbf{1}, the states for which we denote as $\Psi_5$ and $\Psi_1$, respectively.  They lead to the mass terms
\begin{equation}\label{eqA:lag}
\mathcal{L} \supset y_L f \overline{q_L} U \Psi_R + y_R f \overline{\Psi_L} U u_R 
- m_5\overline{\Psi}_{5L}\Psi_{5R}- m_1\overline{\Psi}_{1L}\Psi_{1R} + h.c.
\end{equation}
The states of the \textbf{6} are
\begin{equation}
\Psi = \frac{1}{\sqrt{2}} \left(\begin{array}{c} iB-iX_{5/3}\\ B+X_{5/3}\\ i X_{2/3} + i T\\ X_{2/3} - T\\ i\sqrt{2} T_{a}\\  \sqrt{2} T_{b}
\end{array}\right),
\end{equation}
where under the SM these are $(T,B)$, $(X_{5/3}, X_{2/3})$, $T_a$, and $T_b$ which are respectively a \textbf{2}$_{7/6}$, a \textbf{2}$_{1/6}$, a \textbf{1}$_{2/3}$, and a \textbf{1}$_{2/3}$.  The upper 5 components comprise $\Psi_5$ and the lowest is $\Psi_1$.

The actual calculation of the effective coupling to the field strengths can be simplified using the Higgs low energy theorem that allows us to compute the contribution using only the mass spectrum.  In particular, for a top partner $\Psi^i$ we need to know the Yukawa coupling $g_i$ and the mass $m_i$ defined as $i g_i \eta \Psi^i \gamma_5 \Psi^i$ and $m_i \Psi^i \Psi^i$. With reference to Eq.~\eqref{eqA:lag} we note that $m_i$ is a function of $h$ and $\eta$, but given the assumption of CP conservation $m_i=m_i(\eta=0)$.   On the other hand, $g_i$ can be computed from the imaginary part of the mass matrix $\mathcal{M}$ in the background of $\eta$, $i g_i = \partial m_i/\partial\eta|_{\eta=0}$. The following relation holds,
\begin{equation}
\sum_{i} \frac{g_i}{m_i} = \frac{\partial}{\partial\eta}\log\det\mathcal{M} = \frac{1}{f}\frac{\cot\theta}{\sqrt{1-v^2/f^2}},
\end{equation}
where $i$ runs over the fermion states including the SM top.  This contribution is equal to the contribution just from the top in Eq.~\eqref{eqA:fermion-couplings} which means that the contribution of fermions much heavier than $\eta$ vanishes.  Notice that differently from the case of the Higgs couplings~\cite{Azatov:2011qy} here the wave function renormalization of the light quarks does not introduce new effects (unless CP is broken).

The overall contribution from top partners is then
\begin{equation}
\frac{c_\gamma}{\Lambda} 
= \frac{4}{3} \sum_{i} \frac{g_i}{m_i} A_{-}(\tau_i)
\simeq \frac{1}{f} \cdot \mathcal{O}\left(\frac{m_t}{m_*} \frac{m_\eta^2}{m_*^2}\right)
\end{equation}
This suggests that one has to deviate from the limit of all heavy top partners, however, as discussed in Sec.~\ref{sec:pseudoscalarmass}, it seems challenging to achieve the size needed for $c_\gamma$ and $c_g$ solely from top partners and comply with the direct limits on their masses.

\pagestyle{plain}
\bibliographystyle{jhep}
\small\bibliography{refs}

\providecommand{\href}[2]{#2}\begingroup\raggedright\begin{thebibliography}{10}

\bibitem{ATLAS-CONF-2015-081}
{\it {Search for resonances decaying to photon pairs in 3.2 fb$^{-1}$ of $pp$
  collisions at $\sqrt{s}$ = 13 TeV with the ATLAS detector}},  Tech. Rep.
  ATLAS-CONF-2015-081, CERN, Geneva, Dec, 2015.

\bibitem{CMS-PAS-EXO-15-004}
{\bf CMS Collaboration} Collaboration, {\it {Search for new physics in high
  mass diphoton events in proton-proton collisions at 13TeV}},  Tech. Rep.
  CMS-PAS-EXO-15-004, CERN, Geneva, 2015.

\bibitem{Aad:2015mna}
{\bf ATLAS} Collaboration, G.~Aad et~al., {\it {Search for high-mass diphoton
  resonances in $pp$ collisions at $\sqrt{s}=8$ TeV with the ATLAS detector}},
  {\em Phys. Rev.} {\bf D92} (2015), no.~3 032004,
  [\href{http://arxiv.org/abs/1504.05511}{{\tt arXiv:1504.05511}}].

\bibitem{CMS:2015cwa}
{\bf CMS} Collaboration, C.~Collaboration, {\it {Search for High-Mass Diphoton
  Resonances in pp Collisions at sqrt(s)=8 TeV with the CMS Detector}}, .

\bibitem{Khachatryan:2015qba}
{\bf CMS} Collaboration, V.~Khachatryan et~al., {\it {Search for diphoton
  resonances in the mass range from 150 to 850 GeV in pp collisions at
  $\sqrt{s} =$ 8 TeV}},  {\em Phys. Lett.} {\bf B750} (2015) 494--519,
  [\href{http://arxiv.org/abs/1506.02301}{{\tt arXiv:1506.02301}}].

\bibitem{slides}
ATLAS and C.~collaborations, {\it {ATLAS and CMS physics results from Run 2}},
  2015.
\newblock
  \href{http://indico.cern.ch/event/442432/}{http://indico.cern.ch/event/442432/}.

\bibitem{Aad:2015fna}
{\bf ATLAS} Collaboration, G.~Aad et~al., {\it {A search for $ t\overline{t} $
  resonances using lepton-plus-jets events in proton-proton collisions at
  $\sqrt{s}=8 $ TeV with the ATLAS detector}},  {\em JHEP} {\bf 08} (2015) 148,
  [\href{http://arxiv.org/abs/1505.07018}{{\tt arXiv:1505.07018}}].

\bibitem{Khachatryan:2015sma}
{\bf CMS} Collaboration, V.~Khachatryan et~al., {\it {Search for Resonant
  $\mathrm{t\bar{t}}$ Production in Proton-Proton Collisions at $\sqrt{s}$ = 8
  TeV}},  \href{http://arxiv.org/abs/1506.03062}{{\tt arXiv:1506.03062}}.

\bibitem{Khachatryan:2015tra}
{\bf CMS} Collaboration, V.~Khachatryan et~al., {\it {Search for Neutral MSSM
  Higgs Bosons Decaying into A Pair of Bottom Quarks}},  {\em JHEP} {\bf 11}
  (2015) 071, [\href{http://arxiv.org/abs/1506.08329}{{\tt arXiv:1506.08329}}].

\bibitem{Aad:2014fha}
{\bf ATLAS} Collaboration, G.~Aad et~al., {\it {Search for new resonances in
  $W\gamma$ and $Z\gamma$ final states in $pp$ collisions at $\sqrt s=8$ TeV
  with the ATLAS detector}},  {\em Phys. Lett.} {\bf B738} (2014) 428--447,
  [\href{http://arxiv.org/abs/1407.8150}{{\tt arXiv:1407.8150}}].

\bibitem{Aad:2015kna}
{\bf ATLAS} Collaboration, G.~Aad et~al., {\it {Search for an additional, heavy
  Higgs boson in the $H\rightarrow ZZ$ decay channel at $\sqrt{s}$ = 8 TeV in
  $pp$ collision data with the ATLAS detector}},
  \href{http://arxiv.org/abs/1507.05930}{{\tt arXiv:1507.05930}}.

\bibitem{Aad:2014xka}
{\bf ATLAS} Collaboration, G.~Aad et~al., {\it {Search for resonant diboson
  production in the $\mathrm {\ell \ell }q\bar{q}$ final state in $pp$
  collisions at $\sqrt{s} = 8$ TeV with the ATLAS detector}},  {\em Eur. Phys.
  J.} {\bf C75} (2015) 69, [\href{http://arxiv.org/abs/1409.6190}{{\tt
  arXiv:1409.6190}}].

\bibitem{Khachatryan:2015cwa}
{\bf CMS} Collaboration, V.~Khachatryan et~al., {\it {Search for a Higgs Boson
  in the Mass Range from 145 to 1000 GeV Decaying to a Pair of W or Z Bosons}},
   {\em JHEP} {\bf 10} (2015) 144, [\href{http://arxiv.org/abs/1504.00936}{{\tt
  arXiv:1504.00936}}].

\bibitem{Khachatryan:2014gha}
{\bf CMS} Collaboration, V.~Khachatryan et~al., {\it {Search for massive
  resonances decaying into pairs of boosted bosons in semi-leptonic final
  states at $\sqrt{s} =$ 8 TeV}},  {\em JHEP} {\bf 08} (2014) 174,
  [\href{http://arxiv.org/abs/1405.3447}{{\tt arXiv:1405.3447}}].

\bibitem{Aad:2015ufa}
{\bf ATLAS} Collaboration, G.~Aad et~al., {\it {Search for production of
  $WW/WZ$ resonances decaying to a lepton, neutrino and jets in $pp$ collisions
  at $\sqrt{s}=8$ TeV with the ATLAS detector}},  {\em Eur. Phys. J.} {\bf C75}
  (2015), no.~5 209, [\href{http://arxiv.org/abs/1503.04677}{{\tt
  arXiv:1503.04677}}]. [Erratum: Eur. Phys. J.C75,370(2015)].

\bibitem{Aad:2014aqa}
{\bf ATLAS} Collaboration, G.~Aad et~al., {\it {Search for new phenomena in the
  dijet mass distribution using $p-p$ collision data at $\sqrt{s}=8$ TeV with
  the ATLAS detector}},  {\em Phys. Rev.} {\bf D91} (2015), no.~5 052007,
  [\href{http://arxiv.org/abs/1407.1376}{{\tt arXiv:1407.1376}}].

\bibitem{CMS-PAS-EXO-14-005}
{\bf CMS Collaboration} Collaboration, {\it {Search for Resonances Decaying to
  Dijet Final States at $\sqrt{s} = 8$ TeV with Scouting Data}},  Tech. Rep.
  CMS-PAS-EXO-14-005, CERN, Geneva, 2015.

\bibitem{Aad:2014cka}
{\bf ATLAS} Collaboration, G.~Aad et~al., {\it {Search for high-mass dilepton
  resonances in pp collisions at $\sqrt{s}=8$  TeV with the ATLAS
  detector}},  {\em Phys. Rev.} {\bf D90} (2014), no.~5 052005,
  [\href{http://arxiv.org/abs/1405.4123}{{\tt arXiv:1405.4123}}].

\bibitem{Khachatryan:2014fba}
{\bf CMS} Collaboration, V.~Khachatryan et~al., {\it {Search for physics beyond
  the standard model in dilepton mass spectra in proton-proton collisions at $
  \sqrt{s}=8 $ TeV}},  {\em JHEP} {\bf 04} (2015) 025,
  [\href{http://arxiv.org/abs/1412.6302}{{\tt arXiv:1412.6302}}].

\bibitem{Aad:2015uka}
{\bf ATLAS} Collaboration, G.~Aad et~al., {\it {Search for Higgs boson pair
  production in the $b\bar{b}b\bar{b}$ final state from pp collisions at
  $\sqrt{s} = 8$ TeVwith the ATLAS detector}},  {\em Eur. Phys. J.} {\bf C75}
  (2015), no.~9 412, [\href{http://arxiv.org/abs/1506.00285}{{\tt
  arXiv:1506.00285}}].

\bibitem{Khachatryan:2015yea}
{\bf CMS} Collaboration, V.~Khachatryan et~al., {\it {Search for resonant pair
  production of Higgs bosons decaying to two bottom quark–antiquark pairs in
  proton–proton collisions at 8 TeV}},  {\em Phys. Lett.} {\bf B749} (2015)
  560--582, [\href{http://arxiv.org/abs/1503.04114}{{\tt arXiv:1503.04114}}].

\bibitem{Khachatryan:2014wca}
{\bf CMS} Collaboration, V.~Khachatryan et~al., {\it {Search for neutral MSSM
  Higgs bosons decaying to a pair of tau leptons in pp collisions}},  {\em
  JHEP} {\bf 10} (2014) 160, [\href{http://arxiv.org/abs/1408.3316}{{\tt
  arXiv:1408.3316}}].

\bibitem{Aad:2014vgg}
{\bf ATLAS} Collaboration, G.~Aad et~al., {\it {Search for neutral Higgs bosons
  of the minimal supersymmetric standard model in pp collisions at $\sqrt{s}$ =
  8 TeV with the ATLAS detector}},  {\em JHEP} {\bf 11} (2014) 056,
  [\href{http://arxiv.org/abs/1409.6064}{{\tt arXiv:1409.6064}}].

\bibitem{Aad:2015wra}
{\bf ATLAS} Collaboration, G.~Aad et~al., {\it {Search for a CP-odd Higgs boson
  decaying to Zh in pp collisions at $\sqrt{s} = 8$ TeV with the ATLAS
  detector}},  {\em Phys. Lett.} {\bf B744} (2015) 163--183,
  [\href{http://arxiv.org/abs/1502.04478}{{\tt arXiv:1502.04478}}].

\bibitem{Khachatryan:2014rra}
{\bf CMS} Collaboration, V.~Khachatryan et~al., {\it {Search for dark matter,
  extra dimensions, and unparticles in monojet events in proton–proton
  collisions at $\sqrt{s} = 8$ TeV}},  {\em Eur. Phys. J.} {\bf C75} (2015),
  no.~5 235, [\href{http://arxiv.org/abs/1408.3583}{{\tt arXiv:1408.3583}}].

\bibitem{Aad:2015zva}
{\bf ATLAS} Collaboration, G.~Aad et~al., {\it {Search for new phenomena in
  final states with an energetic jet and large missing transverse momentum in
  pp collisions at $\sqrt{s}=$8 TeV with the ATLAS detector}},  {\em Eur. Phys.
  J.} {\bf C75} (2015), no.~7 299, [\href{http://arxiv.org/abs/1502.01518}{{\tt
  arXiv:1502.01518}}]. [Erratum: Eur. Phys. J.C75,no.9,408(2015)].

\bibitem{partonlumi}
TWIKI, {\it {SM Higgs production cross sections at 13 - 14 TeV}},  2015.
\newblock
  \href{https://twiki.cern.ch/twiki/bin/view/LHCPhysics/CERNYellowReportPageAt1314TeV}{https://twiki.cern.ch/twiki/bin/view/LHCPhysics/CERNYellowReportPageAt1314TeV}.

\bibitem{Spira:1995mt}
M.~Spira, {\it {HIGLU: A program for the calculation of the total Higgs
  production cross-section at hadron colliders via gluon fusion including QCD
  corrections}},  \href{http://arxiv.org/abs/hep-ph/9510347}{{\tt
  hep-ph/9510347}}.

\bibitem{Heinemeyer:2013tqa}
{\bf LHC Higgs Cross Section Working Group} Collaboration, J.~R. Andersen
  et~al., {\it {Handbook of LHC Higgs Cross Sections: 3. Higgs Properties}},
  \href{http://arxiv.org/abs/1307.1347}{{\tt arXiv:1307.1347}}.

\bibitem{Panico:2015jxa}
G.~Panico and A.~Wulzer, {\it {The Composite Nambu—Goldstone Higgs}},  {\em
  Lect. Notes Phys.} {\bf 913} (2016) pp.--,
  [\href{http://arxiv.org/abs/1506.01961}{{\tt arXiv:1506.01961}}].

\bibitem{Agashe:2004rs}
K.~Agashe, R.~Contino, and A.~Pomarol, {\it {The Minimal composite Higgs
  model}},  {\em Nucl. Phys.} {\bf B719} (2005) 165--187,
  [\href{http://arxiv.org/abs/hep-ph/0412089}{{\tt hep-ph/0412089}}].

\bibitem{Gripaios:2009pe}
B.~Gripaios, A.~Pomarol, F.~Riva, and J.~Serra, {\it {Beyond the Minimal
  Composite Higgs Model}},  {\em JHEP} {\bf 04} (2009) 070,
  [\href{http://arxiv.org/abs/0902.1483}{{\tt arXiv:0902.1483}}].

\bibitem{Serra:2015xfa}
J.~Serra, {\it {Beyond the Minimal Top Partner Decay}},  {\em JHEP} {\bf 09}
  (2015) 176, [\href{http://arxiv.org/abs/1506.05110}{{\tt arXiv:1506.05110}}].

\bibitem{Katz:2005au}
E.~Katz, A.~E. Nelson, and D.~G.~E. Walker, {\it {The Intermediate Higgs}},
  {\em JHEP} {\bf 08} (2005) 074,
  [\href{http://arxiv.org/abs/hep-ph/0504252}{{\tt hep-ph/0504252}}].

\bibitem{Galloway:2010bp}
J.~Galloway, J.~A. Evans, M.~A. Luty, and R.~A. Tacchi, {\it {Minimal Conformal
  Technicolor and Precision Electroweak Tests}},  {\em JHEP} {\bf 10} (2010)
  086, [\href{http://arxiv.org/abs/1001.1361}{{\tt arXiv:1001.1361}}].

\bibitem{Mrazek:2011iu}
J.~Mrazek, A.~Pomarol, R.~Rattazzi, M.~Redi, J.~Serra, and A.~Wulzer, {\it {The
  Other Natural Two Higgs Doublet Model}},  {\em Nucl. Phys.} {\bf B853} (2011)
  1--48, [\href{http://arxiv.org/abs/1105.5403}{{\tt arXiv:1105.5403}}].

\bibitem{ATLAS:2015bea}
{\bf ATLAS} Collaboration, T.~A. collaboration, {\it {Measurements of the Higgs
  boson production and decay rates and coupling strengths using pp collision
  data at √s = 7 and 8 TeV in the ATLAS experiment}}, .

\bibitem{Khachatryan:2014jba}
{\bf CMS} Collaboration, V.~Khachatryan et~al., {\it {Precise determination of
  the mass of the Higgs boson and tests of compatibility of its couplings with
  the standard model predictions using proton collisions at 7 and 8 $\,\text
  {TeV}$}},  {\em Eur. Phys. J.} {\bf C75} (2015), no.~5 212,
  [\href{http://arxiv.org/abs/1412.8662}{{\tt arXiv:1412.8662}}].

\bibitem{Kaplan:1991dc}
D.~B. Kaplan, {\it {Flavor at SSC energies: A New mechanism for dynamically
  generated fermion masses}},  {\em Nucl. Phys.} {\bf B365} (1991) 259--278.

\bibitem{Matsedonskyi:2014mna}
O.~Matsedonskyi, G.~Panico, and A.~Wulzer, {\it {On the Interpretation of Top
  Partners Searches}},  {\em JHEP} {\bf 12} (2014) 097,
  [\href{http://arxiv.org/abs/1409.0100}{{\tt arXiv:1409.0100}}].

\bibitem{Aguilar-Saavedra:2013qpa}
J.~A. Aguilar-Saavedra, R.~Benbrik, S.~Heinemeyer, and M.~Pérez-Victoria, {\it
  {Handbook of vectorlike quarks: Mixing and single production}},  {\em Phys.
  Rev.} {\bf D88} (2013), no.~9 094010,
  [\href{http://arxiv.org/abs/1306.0572}{{\tt arXiv:1306.0572}}].

\bibitem{Contino:2003ve}
R.~Contino, Y.~Nomura, and A.~Pomarol, {\it {Higgs as a holographic
  pseudoGoldstone boson}},  {\em Nucl. Phys.} {\bf B671} (2003) 148--174,
  [\href{http://arxiv.org/abs/hep-ph/0306259}{{\tt hep-ph/0306259}}].

\bibitem{Schmaltz:2004de}
M.~Schmaltz, {\it {The Simplest little Higgs}},  {\em JHEP} {\bf 08} (2004)
  056, [\href{http://arxiv.org/abs/hep-ph/0407143}{{\tt hep-ph/0407143}}].

\bibitem{Hinterbichler:2011tt}
K.~Hinterbichler, {\it {Theoretical Aspects of Massive Gravity}},  {\em Rev.
  Mod. Phys.} {\bf 84} (2012) 671--710,
  [\href{http://arxiv.org/abs/1105.3735}{{\tt arXiv:1105.3735}}].

\bibitem{Han:1998sg}
T.~Han, J.~D. Lykken, and R.-J. Zhang, {\it {On Kaluza-Klein states from large
  extra dimensions}},  {\em Phys. Rev.} {\bf D59} (1999) 105006,
  [\href{http://arxiv.org/abs/hep-ph/9811350}{{\tt hep-ph/9811350}}].

\bibitem{Harigaya:2015ezk}
K.~Harigaya and Y.~Nomura, {\it {Composite Models for the 750 GeV Diphoton
  Excess}},  \href{http://arxiv.org/abs/1512.04850}{{\tt arXiv:1512.04850}}.

\bibitem{Mambrini:2015wyu}
Y.~Mambrini, G.~Arcadi, and A.~Djouadi, {\it {The LHC diphoton resonance and
  dark matter}},  \href{http://arxiv.org/abs/1512.04913}{{\tt
  arXiv:1512.04913}}.

\bibitem{Backovic:2015fnp}
M.~Backovic, A.~Mariotti, and D.~Redigolo, {\it {Di-photon excess illuminates
  Dark Matter}},  \href{http://arxiv.org/abs/1512.04917}{{\tt
  arXiv:1512.04917}}.

\bibitem{Angelescu:2015uiz}
A.~Angelescu, A.~Djouadi, and G.~Moreau, {\it {Scenarii for interpretations of
  the LHC diphoton excess: two Higgs doublets and vector-like quarks and
  leptons}},  \href{http://arxiv.org/abs/1512.04921}{{\tt arXiv:1512.04921}}.

\bibitem{Nakai:2015ptz}
Y.~Nakai, R.~Sato, and K.~Tobioka, {\it {Footprints of New Strong Dynamics via
  Anomaly}},  \href{http://arxiv.org/abs/1512.04924}{{\tt arXiv:1512.04924}}.

\bibitem{Knapen:2015dap}
S.~Knapen, T.~Melia, M.~Papucci, and K.~Zurek, {\it {Rays of light from the
  LHC}},  \href{http://arxiv.org/abs/1512.04928}{{\tt arXiv:1512.04928}}.

\bibitem{Buttazzo:2015txu}
D.~Buttazzo, A.~Greljo, and D.~Marzocca, {\it {Knocking on New Physics' door
  with a Scalar Resonance}},  \href{http://arxiv.org/abs/1512.04929}{{\tt
  arXiv:1512.04929}}.

\bibitem{Pilaftsis:2015ycr}
A.~Pilaftsis, {\it {Diphoton Signatures from Heavy Axion Decays at LHC}},
  \href{http://arxiv.org/abs/1512.04931}{{\tt arXiv:1512.04931}}.

\bibitem{Franceschini:2015kwy}
R.~Franceschini, G.~F. Giudice, J.~F. Kamenik, M.~McCullough, A.~Pomarol,
  R.~Rattazzi, M.~Redi, F.~Riva, A.~Strumia, and R.~Torre, {\it {What is the
  gamma gamma resonance at 750 GeV?}},
  \href{http://arxiv.org/abs/1512.04933}{{\tt arXiv:1512.04933}}.

\bibitem{DiChiara:2015vdm}
S.~Di~Chiara, L.~Marzola, and M.~Raidal, {\it {First interpretation of the 750
  GeV di-photon resonance at the LHC}},
  \href{http://arxiv.org/abs/1512.04939}{{\tt arXiv:1512.04939}}.

\bibitem{Frigerio:2012uc}
M.~Frigerio, A.~Pomarol, F.~Riva, and A.~Urbano, {\it {Composite Scalar Dark
  Matter}},  {\em JHEP} {\bf 07} (2012) 015,
  [\href{http://arxiv.org/abs/1204.2808}{{\tt arXiv:1204.2808}}].

\bibitem{Azatov:2011qy}
A.~Azatov and J.~Galloway, {\it {Light Custodians and Higgs Physics in
  Composite Models}},  {\em Phys. Rev.} {\bf D85} (2012) 055013,
  [\href{http://arxiv.org/abs/1110.5646}{{\tt arXiv:1110.5646}}].

\end{thebibliography}\endgroup
\end{document}